\documentclass{ws-ijmpa-hepph0407200}  

\usepackage{amssymb,amsmath,amsfonts,latexsym,graphicx,mathrsfs}

\newcommand{\thetamass}{\underline{\theta}} 
\newcommand{\deltamass}{\underline{\delta}}
\newcommand{\beq}{\begin{equation}}
\newcommand{\eeq}{\end{equation}}
\newcommand{\beqa}{\begin{eqnarray}}
\newcommand{\eeqa}{\end{eqnarray}}
\newcommand{\no}{\nonumber}

\newcommand {\rhs} {right-hand side}

\newcommand {\CS}  {Chern--Simons}

\newcommand{\SM}   {Standard Model}

\begin{document}

\markboth{Frans R. Klinkhamer}
{Lorentz-Noninvariant Neutrino Oscillations}

%
\catchline{}{}{}{}{}
%

\title{LORENTZ-NONINVARIANT NEUTRINO OSCILLATIONS:\\MODEL AND PREDICTIONS}

\author{FRANS R. KLINKHAMER}

\address{Institute for Theoretical Physics,
University of Karlsruhe (TH), 76128 Karlsruhe, Germany\\
frans.klinkhamer@physik.uni-karlsruhe.de}

\maketitle

\begin{history}
\received{8 December 2004}
\end{history}

\begin{abstract}
We present a three-parameter neutrino-oscillation model for
three flavors of massless neutrinos with
Fermi-point splitting and tri-maximal mixing angles.
One of these parameters is the T--violating phase $\epsilon$,
for which the experimental results from K2K and KamLAND appear
to favor a nonzero value. In this article, we give further model
predictions for neutrino oscillations.
Upcoming experiments will be able to test this simple model
and the general idea of Fermi-point splitting.
Possible implications for proposed experiments and neutrino factories
are also discussed.

\keywords{Nonstandard neutrinos; Lorentz noninvariance; quantum phase transition}
\end{abstract}

\ccode{PACS numbers: 14.60.St, 11.30.Cp, 73.43.Nq}

\section{Introduction}
\label{sec:intro}

Several experiments of the last few years
\cite{SuperK1998}\cdash\cite{SNO2003}
have presented (indirect)
evidence for neutrino oscillations over a travel distance
$L \gtrsim 100\,\mathrm{km}$.
No evidence for neutrino oscillations has been seen at
$L \lesssim 1\,\mathrm{km}$.\cite{CHOOZ2002,PaloVerde2001}

The standard explanation for neutrino oscillations invokes the
mass-difference mechanism; see, e.g.,
Refs.~\refcite{GribovPontecorvo}--\refcite{Lipkin}
for a selection of research papers and
Refs.~\refcite{Bargeretal}--\refcite{McKeownVogel} for two recent reviews.
Another possibility, based on an analogy with condensed-matter
physics,\cite{VolovikBook}
is the Fermi-point-splitting mechanism suggested by Volovik and
the present author.
The idea is that a quantum phase transition could give rise to
Lorentz noninvariance and CPT violation via the splitting of a
multiply degenerate Fermi point; see
Ref.~\refcite{KlinkhamerVolovikJETPL} for an introduction and
Ref.~\refcite{KlinkhamerVolovik} for details.
Regardless of the origin, it may be worthwhile to
study modifications of the neutrino dispersion law other than mass terms,
Fermi-point splitting being one of the simplest such modifications.

A first comparison of this Lorentz-noninvariant mechanism
for neutrino oscillations with the
\emph{current} experimental data from K2K and KamLAND
(with additional input from Super--Kamiokande)
was presented in Ref.~\refcite{KlinkhamerJETPL}.
Rough agreement was found for a two-parameter model with tri-maximal mixing
angles. The comparison with the experimental data indicated
a scale of the order of  $10^{-12}\;{\rm eV}$ for the Fermi-point splitting
of the neutrinos and showed a preference for a nonzero value of the
T--violating phase $\epsilon$.
(Note that this phase $\epsilon$ differs, in general, from the one associated
with mass terms, which will be denoted by $\delta$ in the following.)

In this article, we discuss in detail
a three-parameter model with tri-maximal mixing, which is
a direct extension of the previous two-parameter model.
We, then, try to make specific
predictions for possible \emph{future}\footnote{Speaking
about ``current'' and ``future'' experiments, a reference date of
July 2004 has been used, which corresponds to the original version
of the present article [arXiv:hep-ph/0407200v1].}
experiments,
e.g., MINOS,\cite{MINOS1998,MINOS2003} ICARUS,\cite{ICARUS}
OPERA,\cite{OPERA}
T2K,\cite{JPARCSK2001,H2B2001}  NO$\nu$A,\cite{NOvA}
BNL--$\nu$,\cite{BNL2002} beta beams,\cite{Zucchelli}
neutrino factories,\cite{Geer1997,Blondel2004}
and reactor experiments.\cite{Whitepaper}

Let us emphasize, right from the start, that the model considered
is only one of the simplest possible. It could very well be that
the mixing angles are not tri-maximal and that there are
additional mass-difference effects. However, with both mass
differences and Fermi-point splittings present, there is a
multitude of mixing angles and phases to consider. For this
reason, we study in the present article only the simple
Fermi-point-splitting model with tri-maximal mixing and zero
mass differences, which, at least, makes the predictions
unambiguous. Also, it is not our goal to give the best possible
fit to \emph{all} the existing experimental data
on  neutrino oscillations but, rather, to look for qualitatively new effects
(e.g., T and CPT violation).

The outline of this article is as follows.
In Section \ref{sec:fermi-point-splitting}, we discuss the general
form of the dispersion law with Fermi-point splitting.
In Section \ref{sec:neutrino-model}, we present
a simple three-parameter model for three flavors of massless
neutrinos with Fermi-point splitting and tri-maximal mixing angles.
In Section \ref{sec:oscillation-probabilities}, we calculate the
neutrino-oscillation probabilities for the model considered.
In Section \ref{sec:model-parameters-and-predictions}, we obtain
preliminary values for the three parameters of the  model and make both
general and specific predictions (the latter are
for MINOS and T2K in particular).
In Section \ref{sec:outlook}, finally, we sketch a few scenarios of
what future  neutrino-oscillation experiments could tell us about the
relevance of the Fermi-point-splitting model.

\section{Fermi-Point Splitting}
\label{sec:fermi-point-splitting}

In the limit of vanishing Yukawa couplings,
the  Standard Model fermions are massless Weyl fermions,
which have the following dispersion law:
\begin{eqnarray}
\bigl( E_{a,f}({\bf p}) \bigr)^2  &=&
\Bigl(\,c\, |{\bf p}| + b_{0a}^{(f)} \Bigr)^2 \,,
\label{SMdispLaw-timelike}
\end{eqnarray}
for three-momentum ${\bf p}$, velocity of light $c$, and
parameters $b_{0a}^{(f)} =0$.
Here, $a$ labels the sixteen types of massless left-handed
Weyl fermions in the Standard Model
(with a hypothetical left-handed antineutrino included) and
$f$ distinguishes the three known fermion families.
For right-handed Weyl fermions and the same parameters,
the plus sign inside the square on the \rhs~of Eq.~(\ref{SMdispLaw-timelike})
becomes a minus sign; cf. Refs.~\refcite{KlinkhamerVolovik,CK97}.

The Weyl fermions of the original Stan\-dard Mod\-el have all
$b_{0a}^{(f)}$ in Eq.~(\ref{SMdispLaw-timelike}) vanishing, which makes
for a multiply degenerate Fermi point ${\bf p}={\bf 0}$.
[\,Fermi points (gap nodes)
${\bf p}_n$ are points in three-dimensional momentum space at
which the energy spectrum $E({\bf p})$ of the fermionic
quasi-particle has a zero, i.e., $E({\bf p}_n)=0$.]

Negative pa\-ra\-meters $b_{0a}^{(f)}$ in the dispersion law
(\ref{SMdispLaw-timelike}) split this multiply degenerate Fermi point
into Fermi surfaces.
[\,Fermi surfaces $S_n$ are two-dimensional surfaces in
three-dimensional momentum space on which the energy spectrum $E({\bf p})$
is zero, i.e., $E({\bf p})=0$ for all ${\bf p}  \in S_n$.]
Positive pa\-ra\-meters $b_{0a}^{(f)}$  make the respective Fermi points
disappear altogether, the absolute value of the energy being larger than zero
for all momenta.
See Refs.~\refcite{KlinkhamerVolovikJETPL,KlinkhamerVolovik}
for a discussion of the quantum phase transition
associated with Fermi-point splitting.
Note that the dispersion law (\ref{SMdispLaw-timelike}) with
nonzero parameters $b_{0a}^{(f)}$ selects a class of preferred
reference frames; cf. Refs.~\refcite{CK97}--\refcite{CarrollFieldJackiw}.

Possible effects of Fermi-point splitting include neutrino
oscillations,\cite{KlinkhamerVolovik,KlinkhamerJETPL}
as long as the neutrinos are not too much affected by the mechanism of
mass generation. The hope is that massless (or nearly massless) neutrinos
provide a window on ``new physics,''
with Lorentz invariance perhaps as emergent phenomenon.\cite{VolovikBook}
Let us, then, turn to the
phenomenology of massless neutrinos with Fermi-point splittings.

We start from the observation that the Fermi-point splittings of the
\SM~fermions may have a special pattern tracing back to the
underlying physics, just as happens for
the so-called $\alpha$--phase of spin-triplet pairing in
superconductors.\cite{KlinkhamerVolovikJETPL,KlinkhamerVolovik}
As an example, consider the
following factorized \emph{Ansatz}\cite{KlinkhamerVolovik}
for the timelike splittings of Fermi points:
\begin{equation}
b_{0a}^{(f)} =  Y_a \; b_{0}^{(f)}\,,
\label{SMb0pattern}
\end{equation}
for $a=1,\ldots,16$  and $f= 1,2,3$.
Given the hypercharge values $Y_a$ of the
\SM~fermions, there is only one arbitrary energy scale $b_{0}$ per family.

The motivation of the particular choice (\ref{SMb0pattern}) is that
the induced electromagnetic Chern--Simons-like term
cancels out \emph{exactly}
(the result is directly related to the absence of
perturbative gauge anomalies in the \SM).
This cancellation mechanism allows for $b_{0a}^{(f)}$ values larger than
the experimental upper limit
on the \CS~energy scale,\cite{CarrollFieldJackiw}
which is of the order of $10^{-33}\;{\rm eV}$.
Other radiative effects from nonzero $b_{0a}^{(f)}$  remain to be
calculated; cf.  Ref.~\refcite{MocioiuPospelov}.

The dispersion law of a massless
left-handed neutrino (hypercharge $Y_{\nu_L} = -1$) is now given by
\begin{eqnarray}
\bigl( E_{\nu_L,f}({\bf p}) \bigr)^2  &=&
\Bigl(\,c\, |{\bf p}| - b_0^{(f)}\, \Bigr)^2  \,,
\label{DispLaw-nu}
\end{eqnarray}
with $f=1,2,3,$ for three neutrinos.
The corresponding right-handed antineutrino is taken to have
the following dispersion law:
\begin{eqnarray}
\bigl( E_{\bar\nu_R,f}({\bf p}) \bigr)^2  &=&
\Bigl(\,c\, |{\bf p}| + \sigma\,b_0^{(f)}\, \Bigr)^2  \,,
\label{DispLaw-antinu}
\end{eqnarray}
with $\sigma = \pm 1$.
Starting from Dirac fermions, the natural choice would be $\sigma = -1$;
cf. Refs.~\refcite{KlinkhamerVolovik,CK97}. But, in the absence of a definitive
mechanism for the Fermi-point splitting, we keep both
possibilities open. We do not consider the further generalization of
having $\sigma$ in Eq.~(\ref{DispLaw-antinu}) depend on
the family index $f$.

The dispersion laws
(\ref{DispLaw-nu}) and  (\ref{DispLaw-antinu}) can be considered independently
of the particular splitting pattern for the other \SM~fermions,
of which Eq.~(\ref{SMb0pattern}) is only one example.
The parameters $b_{0}^{(f)}$ in these neutrino dispersion laws
act as chemical potentials. This may be compared
to the role of rest mass in a generalized dispersion law:
\beq
\bigl( E({\bf p}) \bigr)^2
= \Bigl(\,c\, |{\bf p}| \pm b_0\, \Bigr)^2 +  m^2\, c^4
\sim
\left(\,c\, |{\bf p}| \pm b_0 +  \frac{m^2\,c^4}{2\, c\,|{\bf p}|}\,\right)^2 \!,
\label{DispLaw-b0m}
\eeq
for $|{\bf p}|  \gg \max\, \left(|b_0|/c,m c\right)$.
The energy change from a nonzero $b_{0}$ always dominates the effect from
$m c^2$ for large enough momenta $|{\bf p}|$.
In order to search for Fermi-point splitting, it is, therefore,
preferable to use neutrino beams with the highest possible energy.

\section{Neutrino Model}
\label{sec:neutrino-model}

Now define three flavor states $|A\rangle, |B\rangle, |C\rangle$
in terms of the left-handed propagation states $|1\rangle, |2\rangle, |3\rangle$,
which have dispersion
law (\ref{DispLaw-nu}) with parameters $b_0^{(f)}$, $f=1,2,3$.
With a unitary  $3 \times 3$ matrix $U$, the relation between these states is
\begin{eqnarray}
\left( \begin{array}{c}|A\rangle\\ |B\rangle\\ |C\rangle \end{array} \right)
  &=& \; U \;
\left( \begin{array}{c}|1\rangle\\ |2\rangle\\ |3\rangle  \end{array} \right)
   \,.
\label{ABC=U123}
\end{eqnarray}
The matrix $U$ can be parametrized\cite{Bargeretal,McKeownVogel} by
three mixing angles, $\chi_{13}, \chi_{21}, \chi_{32}  \in [0,\pi/2]$,
and one phase, $\epsilon  \in [0,2\pi]$:
\beq
U  \equiv  \left( \begin{array}{ccc}
1 & 0 & 0 \\ 0 & c_{32} & s_{32} \\ 0 & -s_{32} & c_{32}
\end{array} \right)
\cdot
\left( \begin{array}{ccc}
c_{13} & 0 & \;+ s_{13}\,e^{+i\epsilon} \\ 0 & 1 & 0 \\
-s_{13}\,e^{-i\epsilon}& 0 & c_{13}
\end{array} \right)
\cdot
\left( \begin{array}{ccc}
c_{21} & s_{21} & 0 \\ -s_{21}& c_{21} & 0 \\ 0 & 0 & 1
\end{array} \right)   \,,
\label{Vmatrix}
\eeq
using the standard notation $s_{ij}$ and $c_{ij}$ for
$\sin\chi_{ij}$ and $\cos\chi_{ij}$.

In this article, we generalize  the \emph{Ansatz} of Ref.~\refcite{KlinkhamerJETPL}
by allowing for unequal ($r \ne 1$) energy steps
$\Delta b_0 ^{(ji)} \equiv b_0^{(j)}- b_0^{(i)}$
between the first and second families ($i,j=1,2$)
and between the second and third ($i,j=2,3$).
Specifically, we introduce the parameters
\beq
B_0 \equiv \Delta b_0^{(21)}
\equiv b_0^{(2)}- b_0^{(1)} \,, \quad
r   \equiv \Delta b_0^{(32)}/\Delta b_0^{(21)}
\equiv \bigl(\,b_0^{(3)}-b_0^{(2)}\bigr)/B_0\,,
\label{parameters-B0r}
\eeq
which are assumed to be nonnegative.
The overall energy scale of the three
parameters $b_0^{(f)}$ is left unspecified.
Note, however, that positive $b_0^{(f)}$ in Eq.~(\ref{DispLaw-nu}) give rise
to Fermi surfaces, which may or may not be topologically protected;
cf. Refs.~\refcite{VolovikBook,KlinkhamerVolovik}.

The mixing angles are again taken to be tri-maximal,\cite{KlinkhamerJETPL}
\beq
\chi_{13}  = \arctan \sqrt{1/2} \approx \pi/5 \,, \quad
\chi_{21}  = \chi_{32}= \arctan 1=\pi/4 \,.
\label{thetapattern}
\eeq
These particular values maximize, for given phase $\epsilon$, the following
T--violation (CP--non\-con\-ser\-va\-tion) measure:
\beq
J \equiv \frac{1}{8}\, \cos\chi_{13}\,\sin 2\chi_{13}\,
            \sin2\chi_{21}\,\sin 2\chi_{32}\, \sin\epsilon \,,
\label{J}
\eeq
which is independent of phase conventions.\cite{Jarlskog}
For this reason, we prefer to use the values (\ref{thetapattern}) rather
than the \emph{ad hoc} values $\chi_{13}  = \chi_{21}  = \chi_{32}= \pi/4$.

In one of the first articles on neutrino oscillations,
Bilenky and Pontecorvo\cite{BilenkyPontecorvo1976}
emphasized
the essential difference between leptons and quarks,
where the latter have quantum numbers which are conserved by
the strong interactions. They argued that, for two flavors, the only natural
value of the neutrino mixing angle $\chi$ is $0$ or $\pi/4$.
But, at this moment, we do not know of a physical mechanism which
convincingly explains the tri-maximal values (\ref{thetapattern})
and we simply use these values as a working hypothesis.

In short, the model (\ref{ABC=U123})--(\ref{thetapattern}),
for dispersion law (\ref{DispLaw-nu}), has three parameters:
the basic energy-difference scale $B_0 \equiv \Delta b_0 ^{(21)}$,
the ratio $r$ of the consecutive energy
steps $\Delta b_0 ^{(21)}$ and $\Delta b_0 ^{(32)}$,
and the T--violating phase $\epsilon$.
This model will be called the ``simple'' Fermi-point-splitting
model in the following (a more general Fermi-point-splitting model
would have arbitrary mixing angles).

\section{Neutrino-Oscillation Probabilities}
\label{sec:oscillation-probabilities}

The calculation of neutrino-oscillation probabilities is
technically straightforward but conceptually rather subtle;
cf. Refs~\refcite{Kayser}--\refcite{Lipkin}.
It may be the simplest to consider a stationary mono-energetic beam
of left-handed neutrinos originating
from a specified interaction region and propagating over a distance
$L$ to an appropriate detector. That is, the neutrinos start out with a
definite flavor (here, labeled $A,B,C$) and
have a spread in their momenta  (from the finite size of the interaction
region and the Heisenberg uncertainty principle,
$\Delta x_m \;\Delta p_n \geq \delta_{mn}\;\hbar/2\,$).

For dispersion law (\ref{DispLaw-nu}) and large enough
neutrino energy ($E_\nu \geq \mathrm{max}\;[\,b_0^{(f)}\,]\,$),
the tri-maximal model  (\ref{ABC=U123})--(\ref{thetapattern})
gives the following neutrino-oscillation probabilities:
\begin{subequations}\label{probabilities}  
\begin{eqnarray}
P(A \rightarrow B, \,\epsilon) &=&
\frac{2}{9}\, \Bigl[\,
  \,\sin^2 \frac{\Delta_{21}}{2} +
\bigl( 1+ {\sqrt{3}}\, c_{\epsilon} \bigr)
\,\sin^2  \frac{\Delta_{31}}{2} +
         \bigl( 1-  {\sqrt{3}}\, c_{\epsilon} \bigr)
\,\sin^2 \frac{\Delta_{32}}{2}
\nonumber\\&&
 -  (1/2)\,\sqrt{3}\;s_{\epsilon}\,
 \bigl(\sin{\Delta_{21}}-\sin{\Delta_{31}}+\sin{\Delta_{32}}\bigr)
     \, \Bigr] \,,
\label{probabilities-a}\\
P(A \rightarrow C, \,\epsilon) &=&
 \frac{2}{9}\, \Bigl[\, \,\sin^2 \frac{\Delta_{21}}{2} +
 \bigl( 1 - {\sqrt{3}}\, c_{\epsilon} \bigr)
\,\sin^2  \frac{\Delta_{31}}{2} +
        \bigl( 1 + {\sqrt{3}}\, c_{\epsilon} \bigr) \,\sin^2 \frac{\Delta_{32}}{2}
\nonumber\\&&
     +  (1/2)\,\sqrt{3}\;s_{\epsilon}\,
 \bigl(\sin{\Delta_{21}}-\sin{\Delta_{31}}+\sin{\Delta_{32}}\bigr)
 \, \Bigr]\,,
\\
P(A \rightarrow A, \,\epsilon) &=&
1-P(A \rightarrow B, \,\epsilon)-P(A \rightarrow C, \,\epsilon)\,,
\\
P(B \rightarrow C, \,\epsilon) &=&
 \frac{2}{9}\, \Bigl[\; (1/2)
 \left( 3\, {s_{\epsilon}}^2 -1  \right) \,\sin^2 \frac{\Delta_{21}}{2} +
 \,\sin^2  \frac{\Delta_{31}}{2} + \,\sin^2 \frac{\Delta_{32}}{2}
\nonumber\\&&
 - (1/2)\,\sqrt{3}\;s_{\epsilon}\,
 \bigl(\sin{\Delta_{21}}-\sin{\Delta_{31}}+\sin{\Delta_{32}}\bigr)
 \, \Bigr] \,,
\\
P(B \rightarrow A, \,\epsilon) &=&
 \frac{2}{9}\, \Bigl[\,
 \,\sin^2 \frac{\Delta_{21}}{2} +
 \bigl( 1 + {\sqrt{3}}\, c_{\epsilon} \bigr)
\,\sin^2  \frac{\Delta_{31}}{2} +
\bigl( 1 - {\sqrt{3}}\, c_{\epsilon} \bigr) \,\sin^2 \frac{\Delta_{32}}{2}
\nonumber\\&&
 +(1/2)\,\sqrt{3}\;s_{\epsilon}\,
 \bigl(\sin{\Delta_{21}}-\sin{\Delta_{31}}+\sin{\Delta_{32}}\bigr)
        \, \Bigr] \,,
\\
P(B \rightarrow B, \,\epsilon) &=&
1-P(B \rightarrow C, \,\epsilon)-P(B \rightarrow A, \,\epsilon) \,,
\end{eqnarray}   
\begin{eqnarray}
P(C \rightarrow A, \,\epsilon) &=&
 \frac{2}{9}\, \Bigl[\,
 \,\sin^2 \frac{\Delta_{21}}{2} +
 \bigl( 1 - {\sqrt{3}}\, c_{\epsilon} \bigr)\,\sin^2  \frac{\Delta_{31}}{2} +
  \bigl( 1 + {\sqrt{3}}\, c_{\epsilon} \bigr) \,\sin^2 \frac{\Delta_{32}}{2}
\nonumber\\&&
 -(1/2)\,\sqrt{3}\;s_{\epsilon}\,
 \bigl(\sin{\Delta_{21}}-\sin{\Delta_{31}}+\sin{\Delta_{32}}\bigr)
        \, \Bigr] \,,
\\
P(C \rightarrow B, \,\epsilon) &=&
 \frac{2}{9}\, \Bigl[\; (1/2)
\left( 3\, {s_{\epsilon}}^2  -1  \right) \,\sin^2 \frac{\Delta_{21}}{2} +
 \,\sin^2  \frac{\Delta_{31}}{2} + \,\sin^2 \frac{\Delta_{32}}{2}
\nonumber\\&&
 +(1/2)\,\sqrt{3}\;s_{\epsilon}\,
 \bigl(\sin{\Delta_{21}}-\sin{\Delta_{31}}+\sin{\Delta_{32}}\bigr)
         \, \Bigr] \,,
\\
P(C \rightarrow C, \,\epsilon) &=&
1-P(C \rightarrow A, \,\epsilon)-P(C \rightarrow B, \,\epsilon) \,,
\label{probabilities-i}
\end{eqnarray}
\end{subequations}
with $s_\epsilon \equiv \sin\epsilon$, $c_\epsilon \equiv \cos\epsilon$, and
\begin{eqnarray}
\Delta_{fg}  &\equiv& \left( b_0^{(f)}-b_0^{(g)} \right) \frac{L}{\hbar\, c}\:.
\label{Deltab0}
\end{eqnarray}
Specifically, the Fermi-point-splitting \emph{Ansatz} has
\beq
\Delta_{21}=  2\pi\, l \,, \quad
\Delta_{32} =  2\pi\,l \, r   \,,\quad
\Delta_{31} =  2\pi\,l \, (1+r)  \,,
\label{Delta213231}
\eeq
in terms of the dimensionless distance
\beq
l \equiv B_0\, L /(h\,c)
\label{ldefinition}
\eeq
and parameters $B_0$ and $r$ from Eq.~(\ref{parameters-B0r}).

The probabilities for right-handed antineutrinos (flavors
$X,Y \in \{A,B,C\}$) are given by
\beq
P(\bar{X} \rightarrow \bar{Y}, \,\epsilon) =
P(X \rightarrow Y, \,\sigma\epsilon) \,,
\label{probabilitiesCP}
\eeq
as follows by changing $\epsilon \rightarrow -\epsilon$ and
$B_0 \rightarrow - \sigma B_0$ in
Eqs.~(\ref{probabilities-a})--(\ref{probabilities-i}),
where the latter transformation corresponds to
the change from (\ref{DispLaw-nu}) to (\ref{DispLaw-antinu})
with $\sigma=\pm 1$.
Note that the survival probability of right-handed antineutrinos
equals that of left-handed neutrinos,
\begin{eqnarray}
P(\bar{X} \rightarrow \bar{X}, \,\epsilon) &=& P(X \rightarrow X, \,\epsilon) \,,
\label{survivalprobXbar}
\end{eqnarray}
independent of the choice of $\sigma$ in the
antineutrino dispersion law (\ref{DispLaw-antinu}).

The time-reversal asymmetry parameter for oscillation probabilities
between $X$--type and $Y$--type neutrinos ($X \neq Y$) is defined by
\begin{eqnarray}
a^\mathrm{(T)}_{XY} \equiv
\frac{P(Y\rightarrow X, \,\epsilon)-P(X\rightarrow Y, \,\epsilon)}
     {P(Y\rightarrow X, \,\epsilon)+P(X\rightarrow Y, \,\epsilon)}\,.
\label{asymmT}
\end{eqnarray}
For model probabilities (\ref{probabilities}), this asymmetry
parameter is proportional to $\sin\epsilon$.
The corresponding CP asymmetry parameter is given by
\begin{eqnarray}
a^\mathrm{(CP)}_{XY} \equiv
\frac{P(\bar{X}\rightarrow \bar{Y}, \,\epsilon)-P(X\rightarrow Y, \,\epsilon)}
     {P(\bar{X}\rightarrow \bar{Y}, \,\epsilon)+P(X\rightarrow Y, \,\epsilon)}\,.
\label{asymmCP}
\end{eqnarray}
For model probabilities (\ref{probabilities}) and (\ref{probabilitiesCP}),
this last asymmetry parameter
vanishes for $\sigma=+1$ and equals $a^\mathrm{(T)}_{XY}$ for $\sigma=-1$.
In other words, the oscillation
probabilities (\ref{probabilities}) and (\ref{probabilitiesCP})
of the neutrino model
with $\sin\epsilon \neq 0$ violate T and CPT  for the case of $\sigma=+1$
and violate T and CP (but keep CPT invariance intact)
for the case of $\sigma=-1$.
In the absence of a definitive theory, it is up to experiment to
decide between the options $\sigma=\pm 1$.

For completeness, we recall
that the standard mass-difference mechanism gives the following
oscillation probabilities\cite{Bargeretal,McKeownVogel} for
fixed ratio $L/E_\nu$ and small enough mass-square difference
$|\Delta m^2_{21}| \equiv |m^2_2 - m^2_1|$:
\begin{subequations}
\label{Pmass-oscill-atmos}
\begin{eqnarray}
&&P_\mathrm{mass-oscill}(\nu_\mu \rightarrow \nu_e)
\sim
\sin^2 (2\,\thetamass_{\,13}) \, \sin^2 \thetamass_{\,32}  \,
\sin^2 \Omega_{32}\,,
\label{Pmass-oscill-atmosMuE}\\[2mm]
&&P_\mathrm{mass-oscill}(\nu_\mu \rightarrow \nu_\tau)
\sim
\cos^4 \thetamass_{\,13} \, \sin^2 (2\,\thetamass_{\,32})  \,
\sin^2 \Omega_{32}\,,
\label{Pmass-oscill-atmosMuTau}
\end{eqnarray}
\end{subequations}
with $\Omega_{ij} \equiv \Delta m^2_{ij}\,c^4 \,L\,/\,(4\,E_\nu \,\hbar\,c)$.
The mixing angles $\thetamass_{\,ij}$ are associated with
the mass terms in the action and are, in general, different from those
defined by Eqs.~(\ref{ABC=U123})--(\ref{Vmatrix});
cf. Ref.~\refcite{ColemanGlashow}.
In fact, the underlined parameters refer to the pure mass-difference
model with all  Fermi-point-splittings strictly zero.
Observe that the T--violating phase $\deltamass$
has dropped out in Eqs.~(\ref{Pmass-oscill-atmos}ab)
because of the limit $\Delta m^2_{21} \to 0$.

For $|\Delta m^2_{32}|  \gg |\Delta m^2_{21}|$ and considering $L$
to be an average distance, one has effectively \beqa
&&P_\mathrm{mass-oscill}(\nu_e \rightarrow \nu_e)
=P_\mathrm{mass-oscill}(\bar{\nu}_e \rightarrow \bar{\nu}_e) \no\\
&&\sim  1 - (1/2) \,\sin^2 (2\,\thetamass_{\,13}) \,
 - \cos^4 \thetamass_{\,13} \,
 \sin^2 (2\,\thetamass_{\,21}) \, \sin^2 \Omega_{21}\,,
\label{Pmass-oscill-solar}
\eeqa
with the $\Omega_{13}$ and $\Omega_{32}$ terms averaged over.

The expressions (\ref{Pmass-oscill-atmos}ab)
have been used to
analyze the data from K2K\cite{K2K2002,K2K2004} and the expression
(\ref{Pmass-oscill-solar}) for the data from
KamLAND,\cite{KamLAND2002,KamLAND2004} all three theoretical
expressions being independent of the phase $\deltamass$.
In the next section, we will use instead the expressions
(\ref{probabilities})--(\ref{survivalprobXbar})
from the simple Fermi-point-splitting model,
which \emph{are} dependent on the T--violating phase $\epsilon$.
But, for brevity, we will drop the explicit $\epsilon$ dependence,
just writing $P(X \rightarrow Y)$ and $P(\bar{X} \rightarrow \bar{Y})$.

\section{Model Parameters and Predictions}
\label{sec:model-parameters-and-predictions}

In order to make unambiguous predictions, we assume in the main part of
this article that the neutrino mass differences are strictly zero.
It could, however, be that
Eqs.~(\ref{DispLaw-nu})--(\ref{DispLaw-antinu}) are modified by having
(small) mass-square terms on their right-hand-sides,
as in Eq.~(\ref{DispLaw-b0m}), which possibility will be
briefly discussed in Sec.~\ref{sec:outlook}.

\subsection{General model predictions}
\label{sec:general-predictions}

The most general prediction of the Fermi-point-splitting
mechanism of neutrino oscillations is
that the oscillation properties are essentially independent of the
neutrino energies. This implies \emph{undistorted} energy
spectra for the reconstructed $\nu_\mu$ energies in, for example,
the current K2K experiment\cite{K2K2002,K2K2004}
and the future MINOS experiment.\cite{MINOS1998,MINOS2003}
[As argued in Ref.~\refcite{KlinkhamerJETPL},
the energy spectra from K2K\cite{K2K2002}
and KamLAND\cite{KamLAND2002,KamLAND2004} are certainly suggestive of
the mass-difference mechanism but do not rule out
the Fermi-point-splitting mechanism yet.
The same conclusion holds for the zenith-angle distributions of
upward stopping or through-going muons from Super--Kamiokande (SK);
cf. Fig.~39 of Ref.~\refcite{KajitaTotsuka2001},
which appears to have a relatively large uncertainty
on the calculated flux of upward through-going muons.]

Assuming that the energy splittings (\ref{Deltab0}) are responsible for the
neutrino-oscillation results of SK, K2K, and KamLAND (see below),
another prediction\cite{KlinkhamerJETPL} is that
\emph{any} reactor experiment at $L = \ell \approx 1\,\mathrm{km}$
(cf. Refs.~\refcite{CHOOZ2002,PaloVerde2001,Whitepaper}) will have survival
probabilities close to 1, at least up to an accuracy of order
$(\Delta b_0\, \ell/\hbar c)^2 \approx   10^{-4}$
for $|\Delta b_0|  \approx 2 \times 10^{-12}\;{\rm eV}$.

Both predictions hold, of course, only if mass-difference effects
from the generalized dispersion law (\ref{DispLaw-b0m}) are
negligible compared to Fermi-point-splitting effects with
$|\Delta b_0|  = \mathrm{O}(10^{-12}\;{\rm eV})$. This corresponds to
$|\Delta m^2|  \lesssim 10^{-6}\,\mathrm{eV}^2/c^4$ for
(anti)neutrino energies in the $\mathrm{MeV}$ range. For larger
mass differences, the corresponding mixing angles may need to be
small, for example $\theta_{\,13} \lesssim 0.2$ for
$|\Delta m^2_{13}| =  2\times 10^{-3}\;{\rm eV}^2 / c^4$; see
Refs.~\refcite{CHOOZ2002,PaloVerde2001}. As mentioned in the preamble of
this section, our model predictions will be for $\Delta m^2_{ij}
=0$ and $\Delta b_0 ^{(ij)}\ne 0$, with mixing angles $\chi_{ij}$ defined by
Eqs.~(\ref{ABC=U123}), (\ref{Vmatrix}) and (\ref{thetapattern}).

\subsection{Preliminary parameter values}
\label{sec:preliminary-parameter-values}

We now present model probabilities at three dimensionless distances $l$
which have ratios 250:180:735 corresponding to the baselines of the K2K,
KamLAND and MINOS experiments, but, initially,
we focus on the first two distances.
[The dimensionless distance $l$ has been defined in Eq.~(\ref{ldefinition}).]
In Table~\ref{tablePratioscan},
model probabilities are calculated starting from an appropriate dimensionless
distance $\tilde{l}$, for a fixed phase  $\epsilon=\pi/4$ and
different values of the parameter $r$ as defined by Eq.~(\ref{parameters-B0r}).
The $\tilde{l}$ values are chosen so that the following inequality holds:
$P(C\rightarrow C)  \geq 65\,\%$ for $0 \leq l \leq \tilde{l}$.

Fixing the ratio $r$
to the value $1/2$, Table~\ref{tablePdeltascan}
gives model probabilities for different values of the phase  $\epsilon$.
The $\epsilon$ range can be restricted to $(-\pi/2,\pi/2]$ without loss of
generality, because the probabilities
(\ref{probabilities})
are invariant under the combined transformations
$\epsilon \rightarrow \epsilon +\pi$ and $(A,B,C) \rightarrow (A,C,B)$.

\begin{table}[t]
\tbl{Model probabilities $(P)|^{\epsilon,r}_{l}$ for different values of the
energy-splitting ratio $r$ and dimensionless distance $l$,
with fixed phase $\epsilon=\pi/4$.
The probabilities $P$ are calculated from
Eqs.~(\ref{probabilities-a})--(\ref{probabilities-i})
and are given in percent, with a rounding error of $1$.
The last row gives the experimental results
for $(P_{\nu_\mu \rightarrow \nu_\mu},P_{\nu_\mu \rightarrow \nu_e},
P_{\nu_\mu \rightarrow \nu_\tau})$
at $L=250\; \mathrm{km}$ from K2K\protect\cite{K2K2002,K2K2004} and
for $(P_{\bar{\nu}_e\rightarrow \bar{\nu}_e})$
at $L\approx 180\; \mathrm{km}$ from KamLAND.\protect\cite{KamLAND2002,KamLAND2004}
The relative error on these experimental numbers is of the order of
$10\,\%$. Also shown are  possible results for MINOS\protect\cite{MINOS1998}
from standard mass-difference neutrino oscillations, where
the values for $(P_{\nu_\mu \rightarrow \nu_\mu},
P_{\nu_\mu \rightarrow \nu_e}, P_{\nu_\mu \rightarrow \nu_\tau})$
are calculated from Eqs.~(\ref{Pmass-oscill-atmos}ab)
for $L=735\; \mathrm{km}$, $E_\nu = 3\;{\rm GeV}$,
$|\Delta m^2_{32}|= 2\times 10^{-3}\;{\rm eV}^2 / c^4$,
$\sin^2 (2\,\thetamass_{\,32}) = 1$,
and $\sin^2 (2\,\thetamass_{\,13}) = 0.2\,$.\vspace*{2mm}}
{
\renewcommand{\tabcolsep}{-0.05pc}  
\renewcommand{\arraystretch}{1.5}   
\begin{tabular}{@{}cc||c|c|c@{}}\toprule
$\tilde{r}$&$\tilde{l}$&
$\;\bigl( P_{C\rightarrow C},P_{C\rightarrow A},P_{C\rightarrow B}
 \bigr)\Big|^{\;\frac{\pi}{4},\tilde{r}}_{\;\tilde{l}}\;$&
$\;\bigl( P_{\bar{A}\rightarrow \bar{A}}
 \bigr)\Big|^{\;\frac{\pi}{4},\tilde{r}}_{\;\tilde{l}\times \frac{180}{250}}$&
$\;\bigl( P_{C\rightarrow C},P_{C\rightarrow A},P_{C\rightarrow B}
 \bigr)\Big|^{\;\frac{\pi}{4},\tilde{r}}_{\;\tilde{l}\times \frac{735}{250}}$\\[3mm]
\hline\hline
\,1/8\,&\,0.250\,&(\,75\;,\;6\;,\;19\,)&(\,71\,)&(\,74\;,\;
9\;,\;17\,)\\ 1/8    &  0.320
&(\,65\;,\;7\;,\;28\,)&(\,57\,)&(\,89\;,\; 8\;,\; 3\,)\\
[1mm]
1/4\;&\;\;0.230\;\;&(\,75\;,\;3\;,\;22\,)&(\,72\,)&(\,58\;,\;17\;,\;25\,)\\
1/4  &    0.260
&(\,70\;,\;3\;,\;27\,)&(\,65\,)&(\,64\;,\;23\;,\;13\,)\\ 1/4  &
0.290    &(\,65\;,\;3\;,\;32\,)&(\,59\,)&(\,66\;,\;26\;,\; 8\,)\\
[1mm]
1/2  &  0.190
&(\,75\;,\;1\;,\;24\,)&(\,74\,)&(\,27\;,\;29\;,\;44\,)\\ 1/2  &
0.210  &(\,70\;,\;1\;,\;29\,)&(\,69\,)&(\,26\;,\;43\;,\;31\,)\\
1/2  &  0.230
&(\,65\;,\;1\;,\;34\,)&(\,64\,)&(\,25\;,\;56\;,\;19\,)\\
[1mm]
1    &  0.130  &(\,75\;,\;2\;,\;23\,)&(\,79\,)&(\, 6\;,\;28\;,\;66\,)\\
1    &  0.145  &(\,70\;,\;2\;,\;28\,)&(\,74\,)&(\,
2\;,\;44\;,\;54\,)\\ 1    &  0.160
&(\,65\;,\;2\;,\;33\,)&(\,69\,)&(\, 1\;,\;61\;,\;38\,)\\
[1mm]
2    &  0.080  &(\,74\;,\;6\;,\;20\,)&(\,81\,)&(\, 5\;,\;27\;,\;68\,)\\
2    &  0.095
&(\,65\;,\;7\;,\;28\,)&(\,74\,)&(\,10\;,\;40\;,\;50\,)\\
\hline\hline --&-- &\;K2K:\, (\,70\;,\;$\leq 1$\;,\;29\,)\;
&\;KamLAND:\, (\,66\,)\; &\;MINOS:\, (\,66?\;,\;3?\;,\;31?\,)\\
\botrule
\end{tabular}
\label{tablePratioscan}}
\end{table}

\begin{table}[t]
\tbl{Model probabilities $(P)|^{\epsilon,r}_{l}$ for different values of
the phase $\epsilon$ and dimensionless distance $l$,
with fixed energy-splitting ratio $r=1/2$.
The probabilities $P$ are calculated from
Eqs.~(\ref{probabilities-a})--(\ref{probabilities-i})
and are given in percent, with a rounding error of $1$.
For different values of $\epsilon$ and $l$,  $\{X,Y,Z\}$ stands for a
permutation of the basic flavors $\{A,B,C\}$.
For  $\epsilon= +\pi/2$, the same probabilities hold for
cyclic permutations of  $\{C,A,B\}$, that is, $\{X,Y,Z\}$ can be
$\{C,A,B\}$, $\{A,B,C\}$, or $\{B,C,A\}$.
The last line gives the experimental results from K2K and KamLAND;
see Table~\ref{tablePratioscan} for further details.\vspace*{2mm}}
{
\renewcommand{\tabcolsep}{0.3pc}    
\renewcommand{\arraystretch}{1.5}   
\begin{tabular}{@{}ccl||c|c@{}}\toprule
$\tilde{\epsilon}$&$\tilde{l}$&
$\,\{X,Y,Z\}\,$&
$\,\bigl( P_{X\rightarrow X},P_{X\rightarrow Y},P_{X\rightarrow Z}
 \bigr)\Big|^{\;\tilde{\epsilon},\frac{1}{2}}_{\;\tilde{l}}\,$&
$\,\bigl( P_{\bar{Y}\rightarrow \bar{Y}}\bigr)
 \Big|^{\;\tilde{\epsilon},\frac{1}{2}}_{\;\tilde{l}\times \frac{180}{250}}\,$\\[3mm]
\hline\hline
$-\pi/4$\,&\,0.135\,&\,$\{A,C,B\}$\,&(\,75\;,\; 1\;,\;24\,)&(\,92\,)\\
$-\pi/4$\,&\,0.160\,&\,$\{A,C,B\}$\,&(\,66\;,\; 1\;,\;33\,)&(\,90\,)\\
[1mm]
$0$\,     &\,0.135\,&\,$\{A,C,B\}$\,&(\,75\;,\; 1\;,\;24\,)&(\,96\,)\\
$0$\,     &\,0.160\,&\,$\{A,C,B\}$\,&(\,66\;,\; 1\;,\;33\,)&(\,94\,)\\
[1mm]
$+\pi/4$\,&\,0.190\,&\,$\{C,A,B\}$\,&(\,75\;,\; 1\;,\;24\,)&(\,74\,)\\
$+\pi/4$\,&\,0.230\,&\,$\{C,A,B\}$  &(\,65\;,\; 1\;,\;34\,)&(\,64\,)\\
[1mm]
$\,+\pi/2\,$&\,0.135\,&\,$\{C,A,B\}^\text{cycl.}
$\,&(\,75\;,\; 9\;,\;16\,)&(\,86\,)\\
$\,+\pi/2\,$&\,0.160\,&\,$\{C,A,B\}^\text{cycl.}
$  &(\,66\;,\;11\;,\;23\,)&(\,81\,)\\
\hline\hline
--&--&\phantom{xxxx}--
&\;K2K:\, (\,70\;,\;$\leq 1$\;,\;29\,)\;
&\;KamLAND:\, (\,66\,)\;\\
\botrule
\end{tabular}
\label{tablePdeltascan}}
\end{table}

The last rows of Tables~\ref{tablePratioscan} and \ref{tablePdeltascan}
give the experimental results from K2K\cite{K2K2002,K2K2004} and
KamLAND.\cite{KamLAND2002,KamLAND2004}
The K2K numbers are deduced from the expected total number of
events $80 \pm 6$ and the observed numbers of $\nu_\mu$ and $\nu_e$
events, respectively $56$ and $1$ (background?).
The latest KamLAND  value for the average $\bar{\nu}_e$ survival probability
is $0.658 \pm 0.044\,{\rm (stat)} \pm 0.047\,{\rm (syst)}$.

Regarding the comparison of the model values and the K2K data in
Table~\ref{tablePratioscan}, recall that
the normalized Poisson distribution $p(n;\mu)\equiv \exp(-\mu)\,\mu^n\,/n!$ gives
$p(1;6) \sim 1 \%$ and $p(1;3) \sim 15 \%$.
This suggests that the best value for $r$ is somewhere between $1/8$
and $2$.
Similarly, Table~\ref{tablePdeltascan} shows a preference for a value
of $\epsilon$ around $\pi/4$ (or $5\pi/4$ with trivial relabelings).

The basic energy-difference scale $B_0$ of the
model is determined by identifying the $\tilde{l}$ value of one
particular row in Table~\ref{tablePratioscan} or \ref{tablePdeltascan}
with the K2K baseline of $250\,\mathrm{km}$.
Using Eq.~(\ref{ldefinition}), one has
\beq
B_0 =  h\, c\;\; \tilde{l}/L_\mathrm{K2K}
    =   1.04 \times 10^{-12}\;{\rm eV} \,
         \left(\frac{250\:\mathrm{km}}{L_\mathrm{K2K}}\right)\,
         \left(\frac{\tilde{l}}{0.210} \right)\,.
\label{B0fromK2K}
\eeq
From the numbers in Tables~\ref{tablePratioscan} and \ref{tablePdeltascan},
the \emph{combined} K2K and KamLAND results then give the following
``central values'' for the parameters of the model
(\ref{ABC=U123})--(\ref{thetapattern}):
\beq
B_0      \approx 10^{-12}\;{\rm eV} \,,  \quad
r        \approx 1/2 \,,      \quad
\epsilon \approx \pi/4 \! \pmod{\pi}\,,
\label{parameters}
\eeq
with identifications
\begin{subequations}
\label{ABCidentification}
\begin{eqnarray}
\Bigl( |A\rangle\,, |B\rangle\,, |C\rangle  \Bigr)
  &=&
\left.\Bigl( |\nu_e\rangle \,, |\nu_\tau\rangle\,, |\nu_\mu\rangle \Bigr)\,
\right|^{\:\epsilon \approx \pi/4 } ,
\label{ABCidentificationQuarterPi}
\\[1mm]
\Bigl( |A\rangle\,, |B\rangle\,, |C\rangle \Bigr)
  &=&
\left.\Bigl( |\nu_e\rangle \,, |\nu_\mu\rangle\,, |\nu_\tau\rangle \Bigr)\,
\right|^{\:\epsilon \approx 5\pi/4} .
\label{ABCidentificationFiveQuarterPi}
\end{eqnarray}\end{subequations}
Note that it is quite remarkable that the K2K and KamLAND data
can be fitted at all by such a simple model.

Only interactions can distinguish between the options
(\ref{ABCidentification}ab).  If, for example, the flavor states
$|\bar{A}\rangle$ and $|C\rangle$ appear in the
dominant decay mode $\mu^{-}\to  e^{+}\, \bar{\nu}_e\, \nu_\mu $,
one would have  $\epsilon \approx \pi/4$.
If, on the other hand, the flavor states
$|\bar{A}\rangle$ and $|B\rangle$
appear, one would have $\epsilon \approx 5\pi/4$.
Alternatively, the dominant decay mode $\pi^{+} \to \mu^{+}\,\nu_\mu$
would imply $\epsilon=\pi/4$ if a $C$--type neutrino is found and
$\epsilon=5\pi/4$ for a $B$--type neutrino.
Henceforth, we focus on  the $\epsilon=\pi/4$ case,
but our conclusions are independent of this choice.

\subsection{Specific model predictions}
\label{sec:specific-predictions}

\begin{figure}[t]
\begin{center}
\includegraphics[width=6.cm]{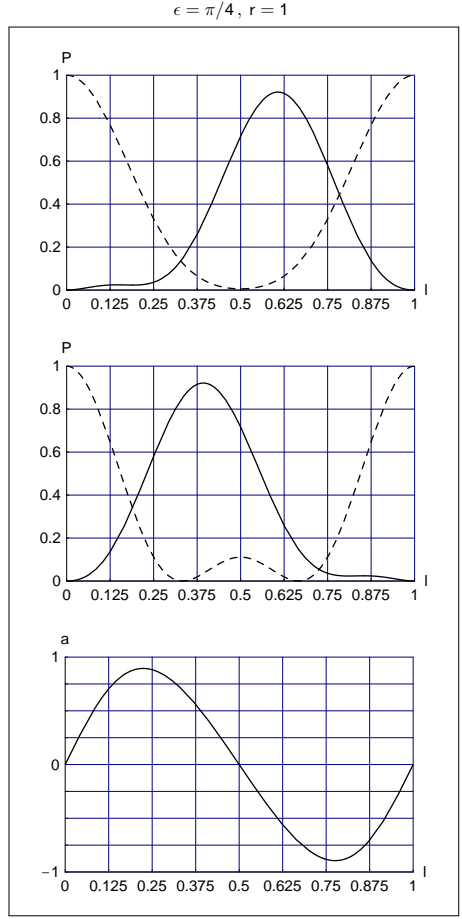}
\end{center}
\vspace*{0cm}
\caption{Model probabilities for vacuum neutrino oscillations
as a function of the dimensionless distance $l$.
The probabilities $P$ are calculated from
Eqs.~(\ref{probabilities-a})--(\ref{probabilities-i}),
with phase $\epsilon=\pi/4$ and energy-splitting ratio $r=1$.
The top panel shows $P(C\rightarrow C)$ [broken curve] and
$P(C\rightarrow A)$ [solid curve], with $P(C\rightarrow B)$ given by
$1-P(C\rightarrow C)-P(C\rightarrow A)$.
The middle panel shows
$P(A\rightarrow A)$ [broken curve] and $P(A\rightarrow C)$ [solid curve],
with $P(A\rightarrow B)=1-P(A\rightarrow A)-P(A\rightarrow C)$.
The bottom panel gives the time-reversal asymmetry (\ref{asymmT}) for
vacuum oscillations between $A$--type and $C$--type neutrinos.
Based on the comparison with the K2K data for flavor identifications
$|A\rangle = |\nu_e\rangle$ and $|C\rangle = |\nu_\mu\rangle$,
the wavelength $l=1$ would correspond to approximately
$1700\,\mathrm{km}$ (cf. Table~\ref{tableLengths}).}
\label{Figratio=one}
\end{figure}

\begin{figure}[t]
\begin{center}
\includegraphics[width=6.cm]{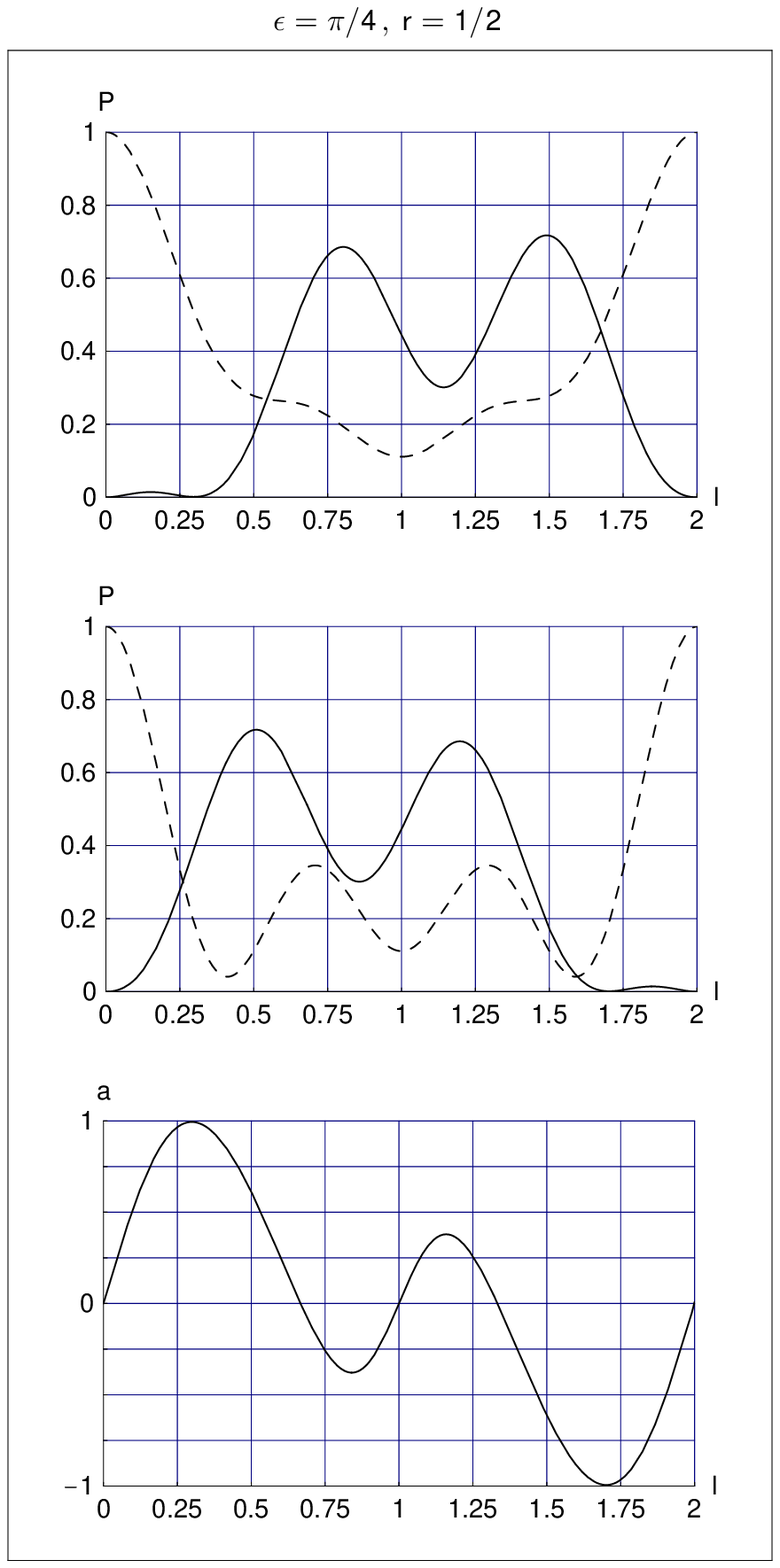}
\end{center}
\vspace*{0cm}
\caption{Same as Fig.~\ref{Figratio=one} but
for the energy-splitting ratio $r=1/2$.
The wavelength $l=2$ would correspond to approximately $2400\,\mathrm{km}$,
according to Table~\ref{tableLengths}.}
\label{Figratio=half}
\end{figure}

\begin{figure}[t]
\begin{center}
\includegraphics[width=6.cm]{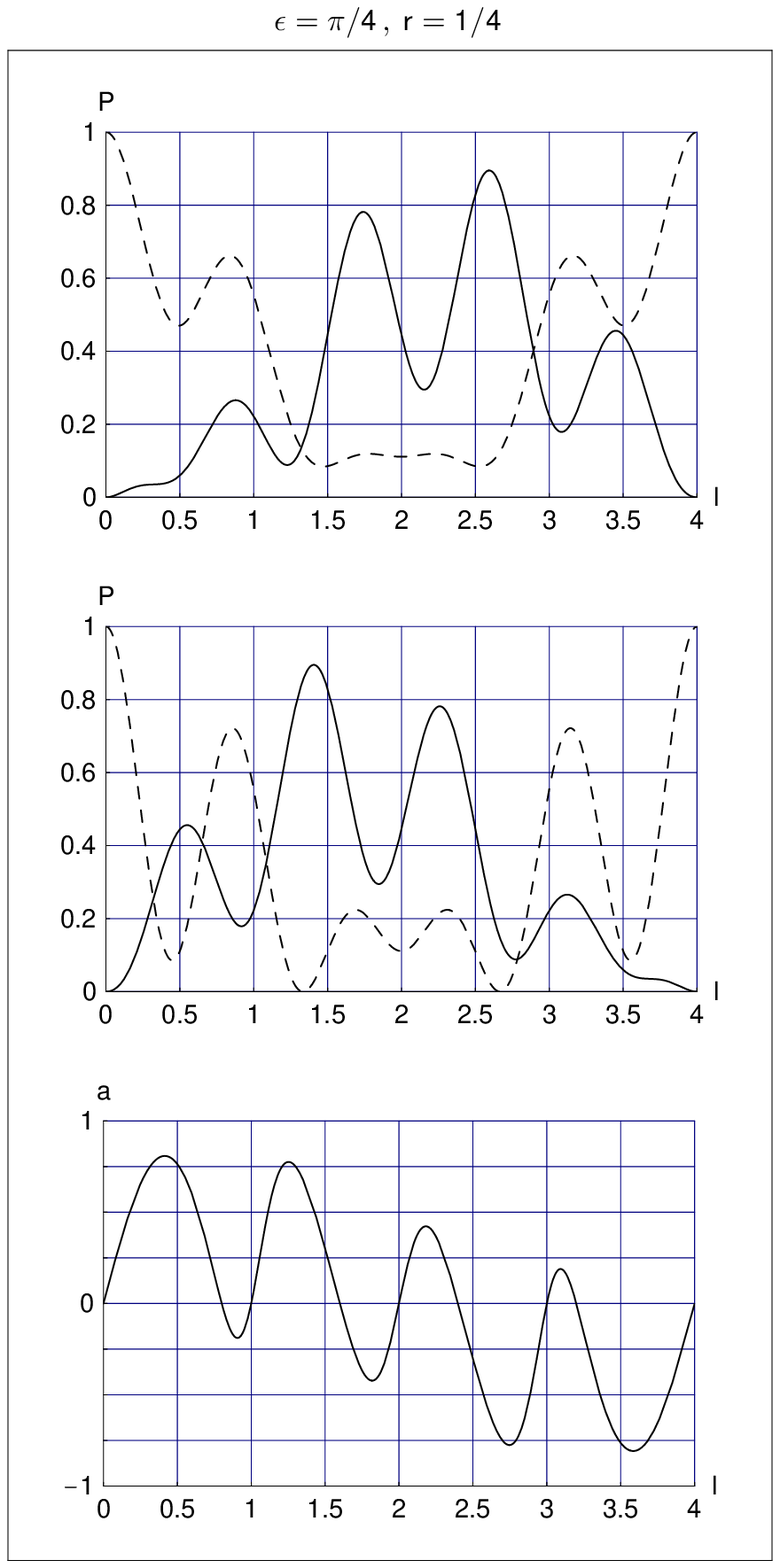}
\end{center}
\vspace*{0cm}
\caption{Same as Fig.~\ref{Figratio=one} but
for the energy-splitting ratio $r=1/4$.
The wavelength $l=4$ would correspond to approximately $3800\,\mathrm{km}$,
according to Table~\ref{tableLengths}.}
\label{Figratio=quarter}
\end{figure}

\begin{figure}[t]
\begin{center}
\includegraphics[width=6.cm]{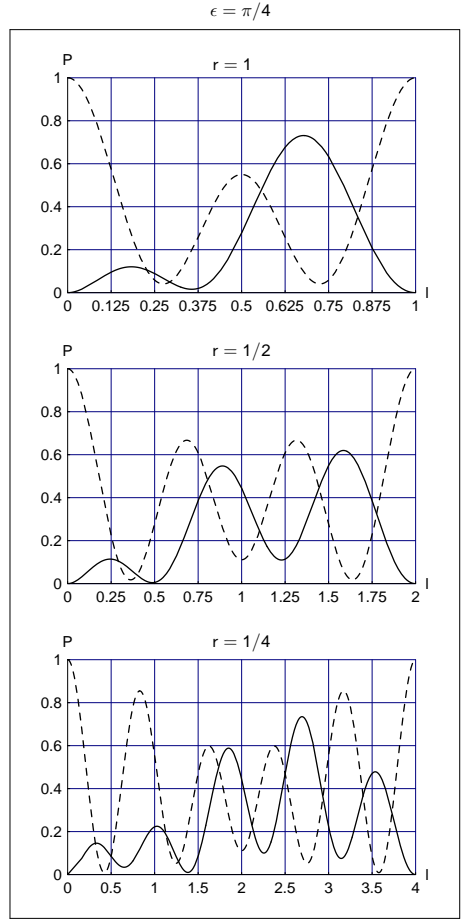}
\end{center}
\vspace*{0cm}
\caption{Model probabilities
$P(B\rightarrow B)$ [broken curve] and $P(B\rightarrow C)$
[solid curve]
as a function of the dimensionless distance $l$,
with fixed phase $\epsilon=\pi/4$ and different energy-splitting ratios $r$.
These probabilities are calculated from
Eqs.~(\ref{probabilities-a})--(\ref{probabilities-i}) and
$P(B\rightarrow A)$ is given by $1-P(B\rightarrow B)-P(B\rightarrow C)$.
The flavor identifications from Eq.~(\ref{ABCidentificationQuarterPi})
would be $|B\rangle =|\nu_\tau\rangle$ and $|C\rangle =|\nu_\mu\rangle$
and, according to Table~\ref{tableLengths},
the wavelengths for $r=1,\,1/2$, and $1/4$ would correspond to
some $1700$, $2400$, and $3800\,\mathrm{km}$, respectively.}
\label{FigProbBtoX}
\end{figure}

In order to prepare for the discussion of new experiments, we give
in Figs.~\ref{Figratio=one}--\ref{Figratio=quarter}
the $\epsilon=\pi/4$  probabilities $P(C\rightarrow X)$  and
$P(A\rightarrow X)$ as a function of the dimensionless travel distance $l$,
for three representative  values of the energy-splitting ratio $r$.
The probabilities $P(B\rightarrow X)$ equal
$1-P(C\rightarrow X)-P(A\rightarrow X)$ and are given
in Fig.~\ref{FigProbBtoX}. These $\epsilon=\pi/4$ probabilities
$P(B\rightarrow X)$ are, of course, of less direct experimental relevance
if $ |B\rangle =|\nu_\tau\rangle$ as suggested by the
identifications (\ref{ABCidentificationQuarterPi}).
Figure~\ref{FigProbBtoX} illustrates rather nicely
the $r$--dependence of the probabilities.
The curves from Fig.~\ref{Figratio=one} and those from the top panel of
Fig.~\ref{FigProbBtoX} are relevant to the two-parameter
($r\equiv 1$) model of Ref.~\refcite{KlinkhamerJETPL}

\begin{figure}[t]
\begin{center}
\includegraphics[width=6.cm]{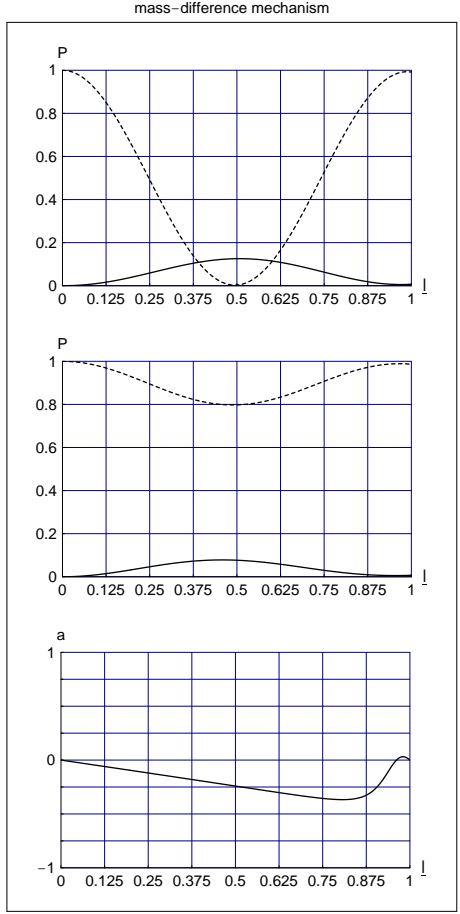}
\end{center}
\vspace*{0cm}
\caption{Probabilities $P$ for vacuum neutrino oscillations due to mass
differences.\protect\cite{Bargeretal,McKeownVogel}
The probabilities
are calculated with the current best values $|\Delta
m^2_{32}|= 2\times 10^{-3}\;{\rm eV}^2 / c^4$, $|\Delta m^2_{21}|=
8 \times 10^{-5}\;{\rm eV}^2 / c^4$, $\sin^2
(2\,\thetamass_{\,32})= 1$, $\sin^2 (2\,\thetamass_{\,21}) = 0.8$,
and maximum allowed values $\sin^2 (2\,\thetamass_{\,13})
= 0.2$, $\deltamass=\pi/2$.
 The top panel shows
$P(\nu_\mu\rightarrow \nu_\mu)$ [broken curve] and $P(\nu_\mu\rightarrow \nu_e)$
[solid curve], as a function of the dimensionless distance
$\underline{l} \equiv L\,|\Delta m^2_{32}| c^4/(2E_\nu h c)
= (|\Delta m^2_{32}|\,c^4)/(2\times 10^{-3}\;{\rm eV}^2)\,
(3\,\mathrm{GeV}/E_\nu)\, L/(3720\,\mathrm{km})$.
The middle panel shows $P(\nu_e\rightarrow \nu_e)$ [broken curve] and
$P(\nu_e\rightarrow \nu_\mu)$ [solid curve].
The bottom panel gives the time-reversal asymmetry
$a^\mathrm{(T)}$ for vacuum oscillations between $e$--type and $\mu$--type
neutrinos, as defined by Eq.~(\ref{asymmT}).
Note that the distance  $\underline{l}=1$
(corresponding to $L \approx 3720\,\mathrm{km}$,
for neutrino energy $E_\nu=3\,\mathrm{GeV}$ and chosen parameters)
is much less than the whole wavelength
$\underline{l}=25$ ($L \approx 10^{5}\,\mathrm{km}$).}
\label{Figmass-smallrange}
\end{figure}

\begin{figure}[t]
\begin{center}
\includegraphics[width=6.cm]{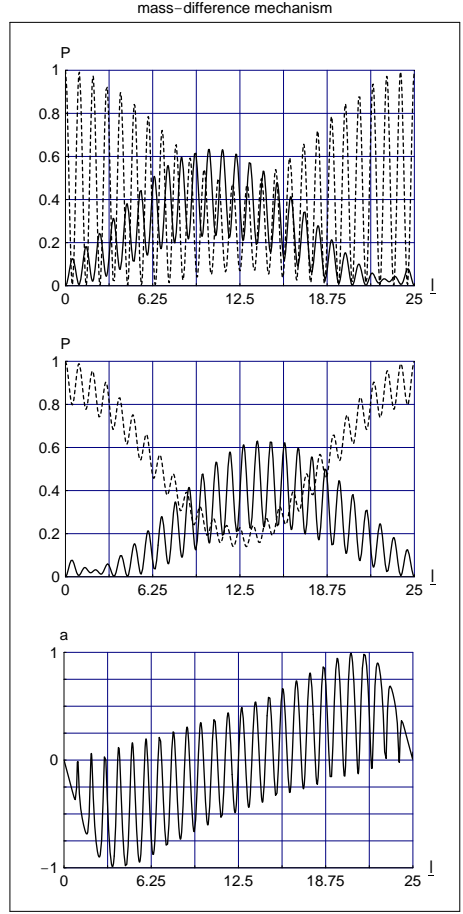}
\end{center}
\vspace*{0cm}
\caption{Same as Fig.~\ref{Figmass-smallrange} but now over the whole
wavelength $\underline{l}=25$ (corresponding to
$L \approx 9.3\times 10^{4}\,\mathrm{km}$, for neutrino
energy $E_\nu=3\,\mathrm{GeV}$ and chosen parameters).}
\label{Figmass-fullrange}
\end{figure}

\begin{figure}[t]
\begin{center}
\includegraphics[width=6.cm]{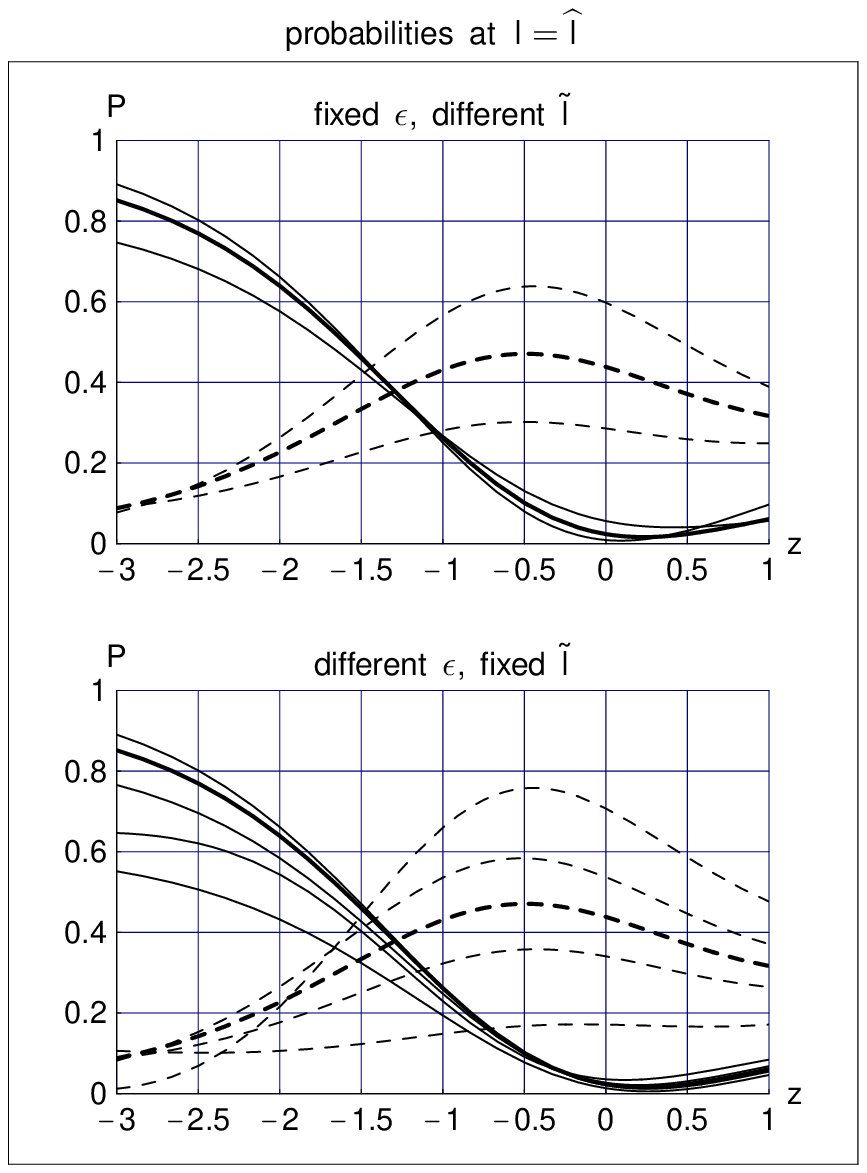}
\end{center}
\vspace*{0cm}
\caption{Model probabilities
$P(\nu_\mu \rightarrow \nu_\mu)$ [solid curves]
and $P(\nu_\mu \rightarrow \nu_e)$  [broken curves]
as a function of parameter $z \equiv \log_{2} r$,
evaluated at the  dimensionless distance
$\widehat{l} \equiv  \tilde{l}\times 735/250$
and with flavor identifications (\ref{ABCidentification}).
The curves of the top panel have
phase $\epsilon=\pi/4$
and $\tilde{l}$ defined as the smallest distance for which
$P(\nu_\mu \rightarrow \nu_\mu) = 0.70 \pm 0.05$
[the central value $0.70$ corresponding to the heavy curves].
Specifically for the top panel,
the broken curves at $z=1$ and
the solid curves at $z=-3$, both from top to bottom, have
$P(\nu_\mu \rightarrow \nu_\mu)[\,\tilde{l}\,]=0.65,\, 0.70,\, 0.75$,
respectively.
The curves of the bottom panel have
length $\tilde{l}$ defined as the smallest distance for which
$P(\nu_\mu \rightarrow \nu_\mu) = 0.70$
and phase $\epsilon=5\pi/32,\, 7\pi/32,\, \pi/4,\, 9\pi/32,\, 11\pi/32$
[the central value $\pi/4$ corresponding to the heavy curves].
Specifically for the bottom panel,
the broken curves at $z=1$ from top to bottom have
$\epsilon\times 32/\pi=5,\, 7,\, 8,\, 9,\, 11$, respectively, and
the solid curves at $z=-3$ from top to bottom have
$\epsilon\times 32/\pi=7,\, 8,\, 9,\, 5,\, 11$, respectively.
The curves of this figure, with $\tilde{l}$ corresponding to
the K2K baseline of $250\,\mathrm{km}$, may be relevant to the
forthcoming MINOS experiment with a baseline of $735\,\mathrm{km}$.
For comparison, the standard mass-difference model gives MINOS
probabilities $P(\nu_\mu \rightarrow \nu_\mu) \approx 66\,\%$
and $P(\nu_\mu \rightarrow \nu_e) \lesssim 3 \,\%$
for $E_\nu \approx 3\;\text{GeV}$ and $\sin^2 (2\thetamass_{13})\leq 0.2\,$;
cf. Table \ref{tablePratioscan}.}
\label{FigProbMINOS}
\end{figure}

\begin{figure}[t]
\begin{center}
\includegraphics[width=6.cm]{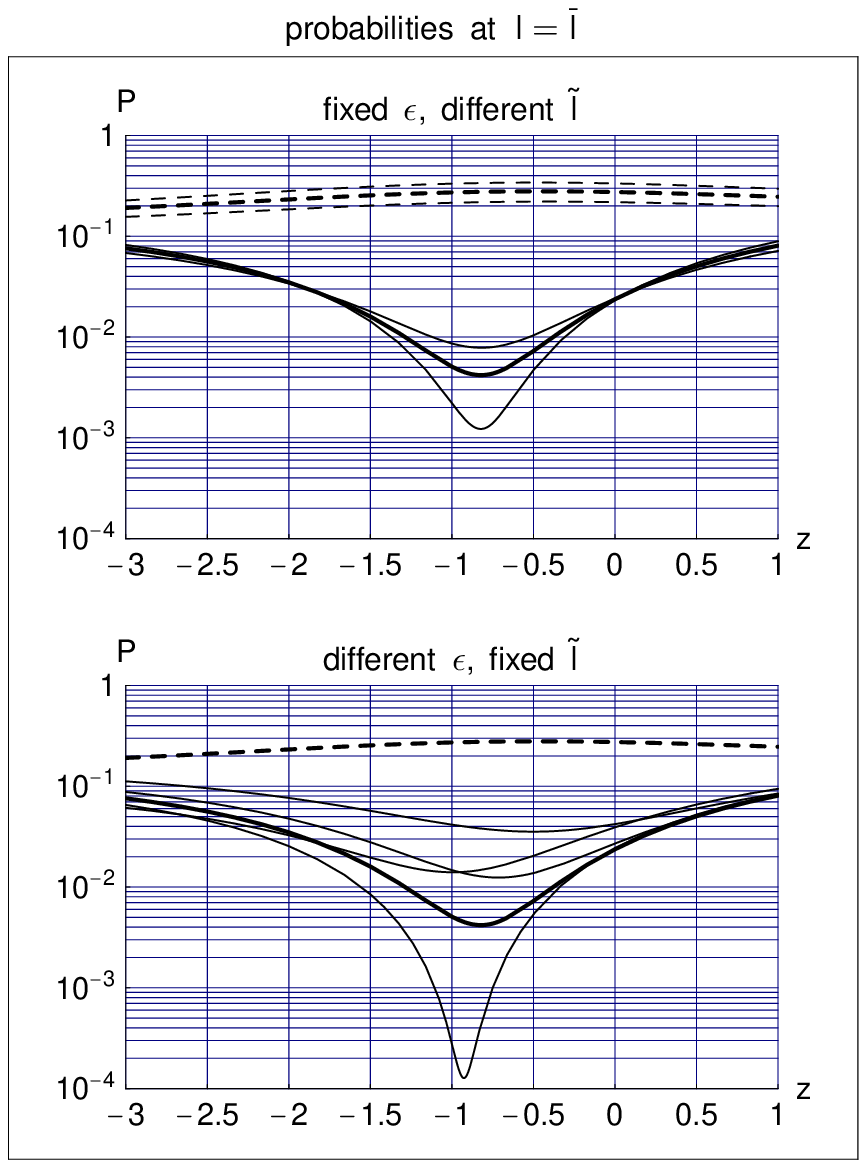}
\end{center}
\vspace*{0cm}
\caption{Model probabilities
$P(\nu_\mu \rightarrow \nu_e)$ [solid curves]
and $P(\nu_e \rightarrow \nu_\mu)$  [broken curves]
as a function of parameter $z \equiv \log_{2} r$,
evaluated at the  dimensionless distance
$\bar{l} \equiv  \tilde{l}\times 295/250$
and with flavor identifications (\ref{ABCidentification}).
The curves of the top panel have
phase $\epsilon=\pi/4$
and $\tilde{l}$ defined as the smallest distance for which
$P(\nu_\mu \rightarrow \nu_\mu) = 0.70 \pm 0.05$
[the central value $0.70$ corresponding to the heavy curves].
Specifically for the top panel,
the solid curves from top to bottom and
the broken curves from bottom to top, both at $z=-1$, have
$P(\nu_\mu \rightarrow \nu_\mu)[\,\tilde{l}\,]=0.75,\, 0.70,\, 0.65$,
respectively.
The curves of the bottom panel have
length $\tilde{l}$ defined as the smallest distance for which
$P(\nu_\mu \rightarrow \nu_\mu) = 0.70$
and phase $\epsilon=5\pi/32,\, 7\pi/32,\,
\pi/4,\, 9\pi/32,\, 11\pi/32$
[the central value $\pi/4$ corresponding to the heavy curves].
Specifically for the bottom panel,
the solid curves at $z=-0.5$ from top to bottom have
$\epsilon\times 32/\pi= 11,\, 5,\, 9,\, 8,\, 7$,
respectively, whereas the broken curves have a rather narrow
band on this logarithmic plot
and only the heavy broken curve for $\epsilon=\pi/4$ is shown.
The solid curves of this figure, with $\tilde{l}$ corresponding to
the K2K baseline of $250\,\mathrm{km}$, may be relevant to the
planned T2K (JPARC--SK) experiment with a baseline of $295\,\mathrm{km}$.
For comparison, the standard mass-difference model gives a T2K
probability  $P(\nu_\mu \rightarrow \nu_e) \lesssim 10 \,\%$,
according to Eq.~(\ref{Pmass-oscill-atmosMuE}) for $\thetamass_{32}=\pi/4$,
$\sin^2 (2\thetamass_{13})\leq 0.2$, and $L$ near
the first oscillation maximum.}
\label{FigProbJPARCSK}
\end{figure}

\begin{figure}[t]
\begin{center}
\includegraphics[width=6.cm]{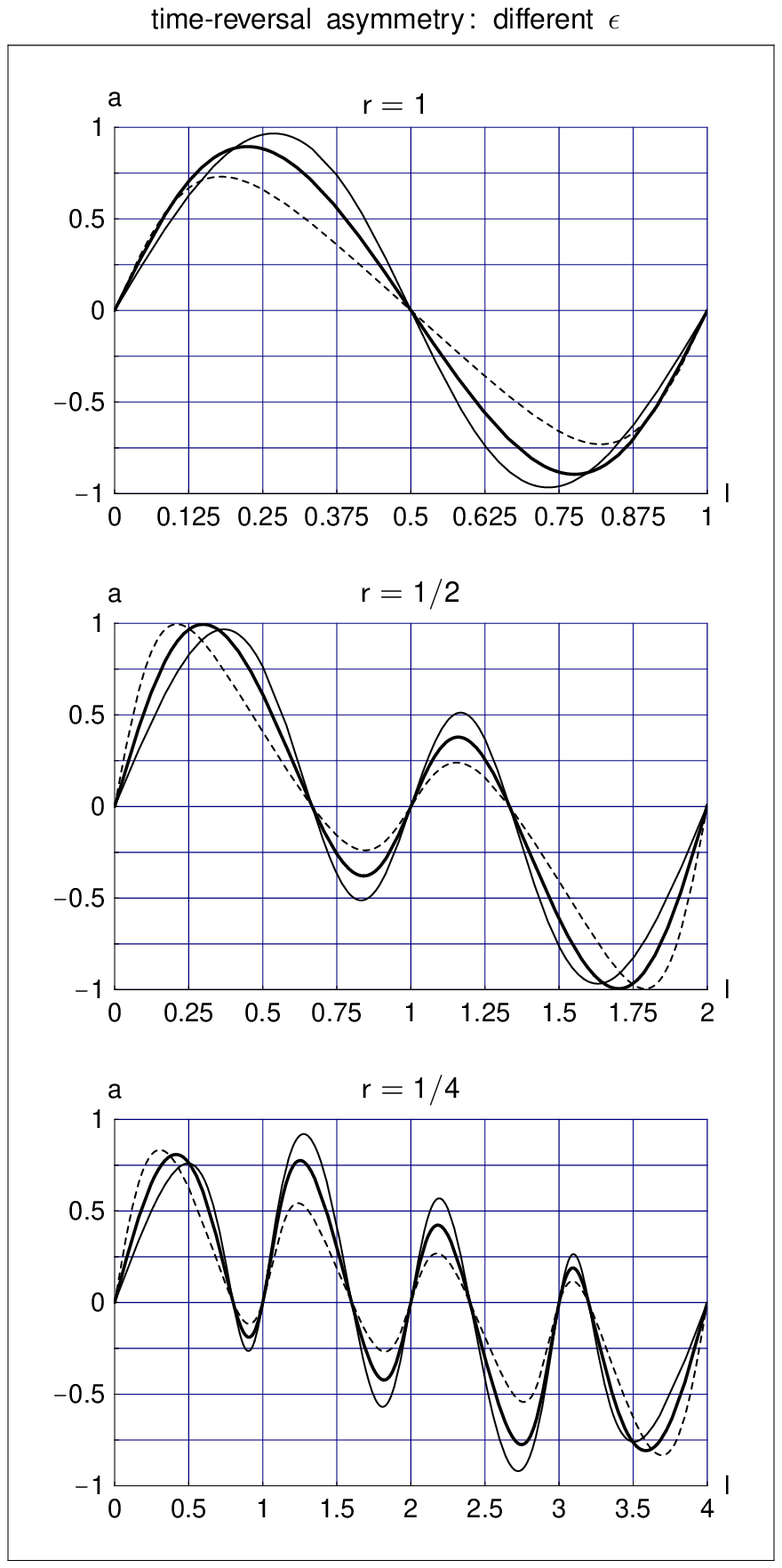}
\end{center}
\vspace*{0.cm}
\caption{Model predictions for the
time-reversal asymmetry (\ref{asymmT}) of vacuum
oscillations between $e$--type and $\mu$--type neutrinos
for different values of the energy-splitting ratio $r$
and different values of the phase,
$\epsilon=5\pi/32,\,  \pi/4,\, 11\pi/32$
[the central value $\pi/4$ corresponding to the heavy curves,
the low value to the broken curves, and the high value to the thin curves].
The probabilities entering the asymmetry parameter
$a$ are calculated from
Eqs.~(\ref{probabilities-a})--(\ref{probabilities-i}) and the flavor
identifications are given by Eqs.~(\ref{ABCidentification}).
According to Table~\ref{tableLengths},
the wavelengths for $r=1,\,1/2$, and $1/4$ would correspond to
some $1700$, $2400$, and $3800\,\mathrm{km}$, respectively.}
\label{FigTasymmdeltas}
\end{figure}

Figures \ref{Figratio=one}--\ref{FigProbBtoX} give the
neutrino-oscillation probabilities for the phase $\epsilon=\pi/4$
but the same curves hold for $\epsilon=5\pi/4$ if the labels
$B$ and $C$ are switched.
Figure \ref{FigProbBtoX}, for example, gives the $\epsilon=5\pi/4$
probabilities $P(C\rightarrow C)$ [broken curve] and
$P(C\rightarrow B)$ [solid curve].
In the same way, Figs.~\ref{Figratio=one}--\ref{Figratio=quarter}
give the $\epsilon=5\pi/4$ probabilities
$P(B\rightarrow B)$ and $P(B\rightarrow A)$ [top panels] and
$P(A\rightarrow A)$ and $P(A\rightarrow B)$ [middle panels].

At this point, we can also mention that the model
probabilities (\ref{probabilities}) for parameters
(\ref{parameters})--(\ref{ABCidentification})
more or less fit the $L/E$ distribution from SK.\cite{SuperK2004}
A rough model estimate for atmospheric
$\mu$--type events (normalized to the expected numbers without neutrino
oscillations) shows, in fact, a significant ``dip'' down to some $45\,\%$
at $L/E$ values just under  $10^3\;\mathrm{km}/\mathrm{GeV}$
and a ``plateau'' at the $58\,\%$ level for larger $L/E$ values,
which more or less agrees  with the experimental data
(cf. Fig.~4 of Ref.~\refcite{SuperK2004}).\footnote{For general
values of $\epsilon$, the model gives an asymptotic value
$(7+\cos 2\epsilon)/12$, assuming an initial ratio
$N_{e-\mathrm{type}}:N_{\mu-\mathrm{type}}:N_{\tau-\mathrm{type}}=1:2:0\,$;
cf. Eqs.~(20) and (26) of Ref.~\refcite{KlinkhamerJETPL}
and Fig.~8 of Ref.~\refcite{KajitaTotsuka2001}. The SK
results\cite{SuperK2004} for the plateau
appear to suggest an $\epsilon$ value away from $0 \pmod \pi$,
just as the KamLAND results\cite{KamLAND2004} in Table~\ref{tablePdeltascan},
but this remains to be confirmed.}
The same calculation gives for the normalized distribution of
atmospheric $e$--type events an average value of $1$ with a
dip at $L/E$ values of a few hundred $\mathrm{km}/\mathrm{GeV}$,
which may perhaps be compared with Fig.~37 of Ref.~\refcite{KajitaTotsuka2001}.
Both dips can be understood heuristically from the relevant curves
in Fig.~\ref{Figratio=half}, but a reliable model calculation
requires further details on the energy spectra
and experimental cuts. Recall that previous studies of atmospheric
neutrino oscillations from alternative models\cite{atmospheric}
were performed in a two-flavor context.

We now turn to the predictions from the simple Fermi-point-splitting
model (\ref{ABC=U123})--(\ref{thetapattern}) for \emph{future} experiments.
As benchmark results, we show in the last row of Table~\ref{tablePratioscan}
probabilities for MINOS from standard mass-difference neutrinos oscillations
(see also Figs.~\ref{Figmass-smallrange} and \ref{Figmass-fullrange}).
Since K2K and MINOS in the low-energy mode
have a similar $L/E$ ratio, the mass-difference mechanism
for neutrino oscillations  (\ref{Pmass-oscill-atmos})
predicts approximately equal probabilities.
For the Fermi-point-splitting mechanism, only the travel distance $L$
enters and different probabilities may be
expected in general. This is born out by the model values shown in the
last column of Table~\ref{tablePratioscan}.

According to Table~\ref{tablePratioscan},
the allowed range of $r$ values from K2K and KamLAND is rather
large, roughly between $1$ to $1/4$. MINOS, on the other hand, would allow
for a more precise determination of $r$, assuming that the simple model has
any validity at all. The top panel of
Fig.~\ref{FigProbMINOS}  gives the model values for
the $\epsilon=\pi/4$ probabilities
$P(\nu_\mu \rightarrow \nu_\mu)$ [solid curves]
and $P(\nu_\mu \rightarrow \nu_e)$ [broken curves]
as a function of $r \equiv 2^{z}$.
These probabilities are evaluated at a
distance $\widehat{l} \equiv  \tilde{l}\times 735/250$
corresponding to the MINOS baseline ($735\; \mathrm{km}$),
provided $\tilde{l}$ corresponds to the K2K baseline ($250\; \mathrm{km}$)
where $P(\nu_\mu \rightarrow \nu_\mu)$
has been measured\cite{K2K2002} to be approximately $0.70 \pm 0.05$.

The simple model already makes the prediction that the appearance probability
$P(\nu_\mu \rightarrow \nu_e)$ from MINOS could be of the order of $10\,\%$
or more (see Fig.~\ref{FigProbMINOS}, top panel).
This is substantially above the expectations from the standard
mass-difference mechanism which predicts $P(\nu_\mu \rightarrow \nu_e)$
of the order of a few percent at most (cf. Fig.~\ref{Figmass-smallrange}),
based on the stringent upper limits for
$\thetamass_{\,13}$ from CHOOZ\cite{CHOOZ2002} and
Palo Verde.\cite{PaloVerde2001}

It is also clear from  Fig.~\ref{FigProbMINOS} [top panel] that a
measurement of the survival probability
$P(\nu_\mu \rightarrow \nu_\mu)$ by MINOS
would indeed allow for a better determination of the value of $r$ in the
model, especially if combined with improved measurements
of the same probability  at $L= 250-295 \;\mathrm{km}$
from K2K\cite{K2K2002,K2K2004} and T2K.\cite{JPARCSK2001}
In fact, the precise determination of $P(\nu_\mu \rightarrow \nu_\mu)$
at $L= 250 \;\mathrm{km}$ would effectively collapse the bands in the
top panel of Fig.~\ref{FigProbMINOS}.
Assuming a measured value of $0.70$
for $P(\nu_\mu \rightarrow \nu_\mu)$ at $L= 250 \;\mathrm{km}$,
the variation with $\epsilon$ around a value of $\pi/4$
is shown in the bottom panel of Fig.~\ref{FigProbMINOS}.
If MINOS would then measure $P(\nu_\mu \rightarrow \nu_\mu)$
between $10 \,\%$ and $60 \,\%$, say, the further measurement of
$P(\nu_\mu \rightarrow \nu_e)$ could be used to constrain the
value of $\epsilon$.
The ICARUS\cite{ICARUS} and OPERA\cite{OPERA} experiments for the CNGS beam
(with a similar baseline of $730\; \mathrm{km}$)  may provide additional
information through the measurement of the other appearance
probability, $P(\nu_\mu \rightarrow \nu_\tau)$, which could be
$10\,\%$ or more according to our model
(see the third entries of the last column of Table~\ref{tablePratioscan}).

Even for the large mixing angle
$\chi_{13}$  of the simple model (\ref{ABC=U123})--(\ref{thetapattern}),
the probability  $P(\nu_\mu \rightarrow \nu_e)$  can be very low
at $l \lesssim 0.25$, but the probability rapidly
becomes substantial for larger values of $l$;
see solid curves in top panels of Figs.~\ref{Figratio=one}--\ref{Figratio=quarter}.
In contrast, mass-difference oscillations (\ref{Pmass-oscill-atmosMuE})
with a rather small $\thetamass_{\,13}$ value\cite{CHOOZ2002,PaloVerde2001}
would have a reduced probability $P(\nu_\mu \rightarrow \nu_e)$
over a larger range of dimensionless distance $\underline{l}\,$;
compare top panels of Figs.~\ref{Figratio=one} and \ref{Figmass-smallrange}.
One of the main objectives of the planned T2K experiment
(previously known as JPARC--SK)
will be to measure this appearance probability at $L= 295 \;\mathrm{km}$.
The relevant model values at $\bar{l} \equiv  \tilde{l}\times 295/250$
are given in Fig.~\ref{FigProbJPARCSK} [solid curves],
together with the values for the time-reversed process [broken curves].
Typical values of $P(\nu_\mu \rightarrow \nu_e)$ are seen to be
around $1\,\%$, but lower values are certainly possible
[for $\epsilon \approx 6.811 \times \pi/32$,
the lowest solid curve of the bottom panel drops to zero at
$z \equiv \log_{2} r \approx -0.9426$].

Another quantity of practical importance to future experiments is the
wavelength $\lambda$  (Table~\ref{tableLengths}).
In addition, there is the length  $L_\mathrm{magic}$
where the time-reversal
asymmetry (\ref{asymmT}) peaks, with $P(\nu_\mu \rightarrow \nu_e)$
significantly less than $P(\nu_e \rightarrow \nu_\mu)$.
Table~\ref{tableLengths} gives the numerical values for this length
and the corresponding asymmetry $a^\mathrm{(T)}_{\mu e}$.
Remarkably, the value of $L_\mathrm{magic}$ is only weakly
dependent on the model parameters $r$ and $\epsilon$;
see Table~\ref{tableLengths} and  Fig.~\ref{FigTasymmdeltas}.
The other magic distance is
$L^\prime\phantom{\!}_\mathrm{magic} \equiv \lambda - L_\mathrm{magic}$,
with $P(\nu_\mu \rightarrow \nu_e)$ significantly larger than
$P(\nu_e \rightarrow \nu_\mu)$. Table~\ref{tableLengths} shows that
the $L^\prime\phantom{\!}_\mathrm{magic}$ value depends strongly on
the model parameter $r$, because $\lambda$ does.
For these large travel distances, matter effects need to be
folded in, but the basic T--asymmetry
$P(\nu_e \rightarrow \nu_\mu)-P(\nu_\mu\rightarrow \nu_e)$
is expected to be unaffected;
cf. Refs.~\refcite{Bargeretal,McKeownVogel,Blondel2004}.

With the identifications (\ref{ABCidentification})
and the large $\chi_{13}$ value (\ref{thetapattern}) of the simple model,
Table~\ref{tableLengths} shows that the distance $L_\mathrm{magic}$
would be at approximately $400\; \mathrm{km}$, which is not very
much more than the T2K baseline of $295\; \mathrm{km}$.\cite{JPARCSK2001}
If the simple model of this article has any relevance,
it would be interesting to have also initial $e$--type neutrinos
for the T2K (JPARC--SK) baseline, possibly from a beta
beam.\cite{Zucchelli,Blondel2004}

\begin{table}[t]
\tbl{Length scales for selected model parameters $B_0$,
$r$, and $\epsilon$.
With these parameters and identifications (\ref{ABCidentification}),
the model (\ref{ABC=U123})--(\ref{thetapattern})
gives  probability $P(\nu_\mu \rightarrow \nu_\mu) \approx 0.70$
at $L= 250\,\mathrm{km}$;
cf. Table~\ref{tablePratioscan} and Eq.~(\ref{B0fromK2K}).
An arbitrary distance $L$ is made dimensionless by defining
$l \equiv B_0\, L /(h\,c)$, in terms of the
basic energy-difference scale $B_0$.
Shown are the (dimensionless) wavelength ($l_\mathrm{wave}$)  $\lambda$,
the (dimensionless) distance ($l_\mathrm{magic}$) $L_\mathrm{magic}$  which
maximizes the time-reversal asymmetry $a^\mathrm{(T)}$ for
$e$--type and $\mu$--type neutrinos as defined by
Eq.~(\ref{asymmT}), and the distance
$L^\prime\protect\phantom{\!}_\mathrm{magic}
\equiv \lambda - L_\mathrm{magic}$
which minimizes the asymmetry (cf. bottom panels
of Figs.~\ref{Figratio=one}--\ref{Figratio=quarter}).\vspace*{2mm}}
{\renewcommand{\tabcolsep}{-0pc}   
\renewcommand{\arraystretch}{1.5}  
\begin{tabular}{@{}ccc||cc|ccc|c@{}}\toprule
$B_0\,[10^{-12}\,\mathrm{eV}]\;$
&$\;r\;$
&$\;\;\epsilon\;$
&$\;\;l_\mathrm{wave}\;$
&$\;\lambda\;[\mathrm{km}]\;$
&$\;\;\;l_\mathrm{magic}\;$
&$\;L_\mathrm{magic}\;[\mathrm{km}]\;$
&$\; a^\mathrm{(T)}_{\mu e}(L_\mathrm{magic})\;$
&$\;\;L^\prime\protect\phantom{\!}_\mathrm{magic}\;[\mathrm{km}]\;$\\[1mm]
\hline\hline
0.72&1          &\,$\;\;\;\pi/4\;\;$ \,&\,1\,
&1724 \,&\;\;\,0.224\,\,&\,386\,&\,$+89\,\%$\,&\,1338\\
1.04&1/2        &\,$\;\;\;\pi/4\;\;$ \,&\,2\,
&2381 \,&\;\;\,0.300\,\,&\,357\,&\,$+99\,\%$\,&\,2024\\
\,1.29\,&\,1/4\,&\,$\;\;\;\pi/4\;\;$ \,&\,4\,
&3846 \,&\;\;\,0.411\,\,&\,395\,&\,$+81\,\%$\,&\,3451\\ \botrule
\end{tabular}\label{tableLengths}}
\end{table}

For model parameters (\ref{parameters}), the middle row of
Table~\ref{tableLengths} shows that the
distance $L^\prime\phantom{\!}_\mathrm{magic}$
would be at approximately $2000\; \mathrm{km}$, which is close
to the JPARC--Beijing baseline of $2100\; \mathrm{km}$ considered in
Ref.~\refcite{H2B2001}. But, as mentioned above, the model parameter
$r$ needs to be determined first, in order to be sure of the
value of $L^\prime\phantom{\!}_\mathrm{magic}$.

\section{Outlook}
\label{sec:outlook}

In this article, we have presented exploratory results for
neutrino-oscillation effects from Fermi-point splitting.
As discussed in the Introduction, the main reason to focus on
the ``simple'' model (\ref{ABC=U123})--(\ref{thetapattern})
for massless neutrinos is the
relatively small number of parameters (essentially three).
With both mass differences and Fermi-point splittings present,
there would be many more mixing angles and phases to consider;
cf. Ref.~\refcite{ColemanGlashow}

As an alternative to the simple (and radical) model with tri-maximal
mixing and zero mass differences, we can briefly mention the
following ``stealth scenario.'' Take neutrino
masses $m_i$ and  mixing angles $\theta_{\,ij}$
(the phase $\delta$ still being arbitrary)
to match approximately the known experimental
results,\cite{Bargeretal,McKeownVogel} leaving out LSND.
Then add Fermi-point splittings with
$|\Delta b_0| \lesssim 10^{-12}\,\mathrm{eV}$,
mixing angles $\chi_{ij} \sim \pi/4$,
and phase $\epsilon \sim \pi/4$.
These additional $b_0$ terms in the dispersion laws  are
not rigorously excluded by \emph{current} experiments,
as the Fermi-point splittings alone can already more or less reproduce
the experimental data.
The Fermi-point-splitting terms could, however, give rise to
new effects for \emph{future} experiments
(e.g., MINOS and T2K), as shown qualitatively by
Figs.~\ref{FigProbMINOS}--\ref{FigTasymmdeltas}.
In other words, there could be surprises around the corner.

Based on the results of this article and the above remarks, one possible
roadmap for nonstandard neutrino oscillations might be the
following:
\vspace*{.5\baselineskip}\newline
\emph{1. Does K2K (or MINOS in the low-energy mode) find little
distortion of the $\nu_\mu$ energy spectrum or, better,
does MINOS find the survival probabilities
$P(\nu_\mu \rightarrow \nu_\mu)$ for the low-energy
beam $\mathit{(\,\langle E_{\nu_\mu} \rangle \approx 3\;GeV)}$
and the high-energy beam
$\mathit{(\,\langle E_{\nu_\mu} \rangle \approx 15\;GeV)}$
to be more or less equal?}
\newline\underline{If so}, go to 2.
\newline\underline{If not},
mass-difference effects may be more important for neutrino oscillations
than Fermi-point-splitting effects, at least for the energies considered.
\vspace*{.5\baselineskip}\newline
\emph{2. Does MINOS in the medium- or high-energy mode
find an appearance probability
$P(\nu_\mu \rightarrow \nu_e)$ above a few percent?}
\newline\underline{If so}, go to 3.
\newline\underline{If not},
the simple Fermi-point-splitting model (\ref{ABC=U123})--(\ref{thetapattern})
would appear to be inappropriate and $\chi_{13}$ might be relatively small
(if the Fermi-point-splitting model with nonmaximal mixing angles
does fit the data, go also to 3).
\vspace*{.5\baselineskip}\newline
\emph{3. Does the (simple) Fermi-point-splitting model fit the data
from K2K, MINOS, and ICARUS/OPERA
with consistent values for $B_0$,
$r$, and $\epsilon\,$?}
\newline\underline{If so}, reconsider the
future options based on the relevant \emph{energy-independent}
length scales of the Fermi-point-splitting model
(cf. Table~\ref{tableLengths} and Fig.~\ref{FigTasymmdeltas}).
These future options include
superbeams,\cite{JPARCSK2001}\cdash\cite{BNL2002}
beta beams,\cite{Zucchelli,Blondel2004}
and neutrino factories.\cite{Geer1997,Blondel2004}
Reactor experiments\cite{Whitepaper}
may also be important
to constrain or determine possible mass terms in
the generalized dispersion law (\ref{DispLaw-b0m})
and the additional mixing angles and phases.
\newline\underline{If not}, change the model or find an entirely
different neutrino-oscillation mechanism.
\vspace*{.5\baselineskip}\newline
The hope is that the first two questions of this roadmap can be answered
in part by the final results from K2K and the initial results from MINOS.

\vspace*{-0\baselineskip}
\section*{Acknowledgments}

It is a pleasure to thank W.J. Marciano, J.G. Morfin,
and A.Yu. Smirnov for discussion and advice,
several par\-ti\-ci\-pants of NuFact04 (Osaka, July 2004) for useful
remarks, and C. Kaufhold for help with the figures.

\vspace*{-0\baselineskip}


\begin{thebibliography}{00}

\bibitem{SuperK1998}
Y. Fukuda {\it et al.}  [Super--Kamiokande Collaboration],
\emph{Phys.\ Rev.\ Lett.}  {\bf 81}, 1562 (1998), \mbox{hep-ex/9807003}.


\bibitem{SuperK2004}
Y. Ashie \emph{et al.}  [Super-Kamiokande Collaboration],
\emph{Phys.\ Rev.\ Lett.}  {\bf 93}, 101801 (2004), \mbox{hep-ex/0404034}.


\bibitem{KajitaTotsuka2001}
T. Kajita and Y. Totsuka,
\emph{Rev.\ Mod.\ Phys.}  {\bf 73}, 85 (2001).

\bibitem{K2K2002}
M.H. Ahn \emph{et al.}  [K2K Collaboration],
\emph{Phys.\ Rev.\ Lett.}  {\bf 90}, 041801 (2003), \mbox{hep-ex/0212007}.


\bibitem{K2K2004}
M.H. Ahn \emph{et al.}  [K2K Collaboration],
\emph{Phys.\ Rev.\ Lett.}  {\bf 93}, 051801 (2004), \mbox{hep-ex/0402017}.


\bibitem{KamLAND2002}
K. Eguchi \emph{et al.}  [KamLAND Collaboration],
\emph{Phys.\ Rev.\ Lett.} {\bf 90}, 021802 (2003), \mbox{hep-ex/0212021}.


\bibitem{KamLAND2004}
T. Araki \emph{et al.} [KamLAND Collaboration],
\emph{Phys. Rev. Lett.} {\bf 94}, 081801 (2005), \mbox{hep-ex/0406035}.




\bibitem{SNO2002}
Q.R. Ahmad \emph{et al.}  [SNO Collaboration],
\emph{Phys.\ Rev.\ Lett.}  {\bf 89}, 011301 (2002), \mbox{nucl-ex/0204008}.


\bibitem{SNO2003}
S.N. Ahmed \emph{et al.}  [SNO Collaboration],
\emph{Phys.\ Rev.\ Lett.}  {\bf 92}, 181301 (2004), \mbox{nucl-ex/0309004}.


\bibitem{CHOOZ2002}
M. Apollonio \emph{et al.},
\emph{Eur.\ Phys.\ J.\ C}  {\bf 27}, 331 (2003), \mbox{hep-ex/0301017}.


\bibitem{PaloVerde2001}
F. Boehm \emph{et al.},
\emph{Phys.\ Rev.\ D}  {\bf 64}, 112001 (2001), \mbox{hep-ex/0107009}.


\bibitem{GribovPontecorvo}
V.N. Gribov and B.~Pontecorvo,
\emph{Phys.\ Lett.\ B}  {\bf 28}, 493 (1969).


\bibitem{BilenkyPontecorvo1976}
S.M. Bilenky and B. Pontecorvo,
\emph{Phys.\ Lett.\ B} {\bf 61}, 248 (1976).


\bibitem{Kayser}
B. Kayser,
\emph{Phys.\ Rev.\ D} {\bf 24}, 110 (1981).


\bibitem{Stodolsky}
L. Stodolsky,
\emph{Phys.\ Rev.\ D} {\bf 58}, 036006 (1998), \mbox{hep-ph/9802387}.


\bibitem{Lipkin}
H.J. Lipkin,
\emph{Phys.\ Lett.\ B} {\bf 579}, 355 (2004), \mbox{hep-ph/0304187}.


\bibitem{Bargeretal}
V. Barger, D. Marfatia, and K. Whisnant,
\emph{Int.\ J.\ Mod.\ Phys.\ E}  {\bf 12}, 569 (2003), \mbox{hep-ph/0308123}.


\bibitem{McKeownVogel}
R.D. McKeown and P. Vogel,
\emph{Phys.\ Rept.} {\bf 394}, 315 (2004), \mbox{hep-ph/0402025}.

\bibitem{VolovikBook}
G.E. Volovik,
\emph{The Universe in a Helium Droplet} (Clarendon Press, Oxford, 2003).

\bibitem{KlinkhamerVolovikJETPL}
F.R. Klinkhamer and G.E. Volovik,
\emph{JETP Lett.} {\bf 80}, 343 (2004), cond-mat/0407597.


\bibitem{KlinkhamerVolovik}
F.R. Klinkhamer and G.E. Volovik,
\emph{Int. J. Mod. Phys. A} {\bf 20}, 2795 (2005), \mbox{hep-th/0403037}.


\bibitem{KlinkhamerJETPL}
F.R. Klinkhamer,
\emph{JETP Lett.} {\bf 79}, 451 (2004), \mbox{hep-ph/0403285}.

\bibitem{MINOS1998}
P. Adamson  \emph{et al.} [MINOS Collaboration],
\emph{The MINOS Detectors Technical Design Report},
FermiLab report NuMI-L-337, October 1998
[MINOS homepage at http://www-numi.fnal.gov].




\bibitem{MINOS2003}
K. Lang,
in: \emph{Proceedings International Workshop on Astroparticle and
High-Energy Physics (AHEP-2003)}, Valencia, Spain, October 2003,
eds.  M. Hirsch \emph{et al.}, 
JHEP Proceedings AHEP-2003/017.


\bibitem{ICARUS}
P. Aprili \emph{et al.}  [ICARUS Collaboration],
\emph{The ICARUS experiment: A second-generation proton decay experiment and
neutrino observatory at the Gran Sasso laboratory. Addendum:
Cloning of T600 modules to reach the design sensitive mass},
report CERN-SPSC-2002-027, August 2002
[ICARUS homepage  at http://www.aquila.infn.it/icarus].

\bibitem{OPERA}
M. Guler \emph{et al.}  [OPERA Collaboration],
\emph{OPERA: An appearance experiment to search for
$\nu_\mu \leftrightarrow \nu_\tau$  oscillations
in the CNGS beam. Experimental proposal},
report CERN-SPSC-2000-028, July 2000
[OPERA homepage  at http://operaweb.web.cern.ch].



\bibitem{JPARCSK2001}
Y. Itow \emph{et al.} [JHF Neutrino Working Group],
in: \emph{Neutrino Oscillations and Their Origin},
eds. Y. Suzuki \emph{et al.}
(World Scientific, Singapore, 2003), p. 239, \mbox{hep-ex/0106019}
[T2K homepage  at http://neutrino.kek.jp/jhfnu/].


\bibitem{H2B2001}
H.S. Chen \emph{et al.}  [VLBL Study Group H2B-1 Collaboration],
report IHEP-EP-2001-01, \mbox{hep-ph/0104266};
Y.F. Wang  \emph{et al.}  
[VLBL Study Group H2B-4 Collaboration],
\emph{Phys.\ Rev.\ D} {\bf 65}, 073021 (2002), \mbox{hep-ph/0111317};
M. Aoki  \emph{et al.},
\emph{Phys.\ Rev.\ D} {\bf 67}, 093004 (2003), \mbox{hep-ph/0112338}.

\bibitem{NOvA}
D.S. Ayres \emph{et al.}  [NO$\nu$A Collaboration],
``Proposal to build a 30 kiloton off-axis detector to study $\nu_\mu \to \nu_e$
oscillations in the NuMI beamline,''  March 2005, \mbox{hep-ex/0503053}
[NO$\nu$A homepage at http://www-nova.fnal.gov/].



\bibitem{BNL2002}
M.V. Diwan \emph{et al.},
\emph{Phys.\ Rev.\ D} {\bf 68}, 012002 (2003), \mbox{hep-ph/0303081}.


\bibitem{Zucchelli}
P. Zucchelli,
\emph{Phys.\ Lett.\ B} {\bf 532}, 166 (2002).


\bibitem{Geer1997}
S. Geer,
\emph{Phys.\ Rev.\ D} {\bf 57}, 6989 (1998); \emph{ibid.} {\bf 59}, 039903 (E) (1999),
\mbox{hep-ph/9712290}.



\bibitem{Blondel2004}
A. Blondel \emph{et al.},
\emph{ECFA/CERN studies of a European neutrino factory complex},
report CERN-2004-002, April 2004 [Chapter 3 available as \mbox{hep-ph/0210192}].


\bibitem{Whitepaper}
K.~Anderson \emph{et al.},
\emph{White paper report on using nuclear reactors to search for a value of
theta(13)}, report FERMILAB-PUB-04-180, January 2004,
\mbox{hep-ex/0402041}.




\bibitem{CK97}
D. Colladay and V.A. Kosteleck\'y,
\emph{Phys.\ Rev.\ D} {\bf 55}, 6760 (1997), \mbox{hep-ph/9703464}.

\bibitem{ColemanGlashow}
S. Coleman and S.L. Glashow,
\emph{Phys.\ Rev.\ D} {\bf 59}, 116008 (1999), \mbox{hep-ph/9812418}.


\bibitem{CarrollFieldJackiw}
S.M. Carroll, G.B. Field, and R. Jackiw,
\emph{Phys.\ Rev.\ D} {\bf 41}, 1231 (1990).


\bibitem{MocioiuPospelov}
I. Mocioiu and M. Pospelov,
\emph{Phys.\ Lett.\ B} {\bf 534}, 114 (2002), \mbox{hep-ph/0202160}.

\bibitem{Jarlskog}
C. Jarlskog,
\emph{Phys.\ Rev.\ Lett.}  {\bf 55}, 1039 (1985).


\bibitem{atmospheric}
P. Lipari and M. Lusignoli,
\emph{Phys.\ Rev.\ D} {\bf 60}, 013003 (1999), \mbox{hep-ph/9901350};
G.L. Fogli, E. Lisi, A. Marrone, and G. Scioscia,
\emph{Phys.\ Rev.\ D} {\bf 60}, 053006 (1999), \mbox{hep-ph/9904248};
M.C. Gonzalez-Garcia and M. Maltoni,
\emph{Phys.\ Rev.\ D} {\bf 70}, 033010 (2004), \mbox{hep-ph/0404085}.

\end{thebibliography}
\end{document}